\documentclass[journal]{IEEEtran}  \newcommand{\figsza}{0.99} \newcommand{\figszb}{0.5} \newcommand{\figszc}{0.8}

\usepackage[cmex10]{amsmath}
\usepackage{amsfonts}
\usepackage{amssymb}
\usepackage{amsopn}
\usepackage{graphicx}
\usepackage{subfigure}
\usepackage{tabularx,colortbl}
\usepackage{mathdots}

\graphicspath{{fig/}}
\newcommand{\bq}{``}
\newcommand{\sups}[1]{\ensuremath{^{\textrm{#1}}}}
\newcommand{\subs}[1]{\ensuremath{_{\textrm{#1}}}}
\newcommand{\co}[0]{CO\subs{2} }
\newcommand{\bo}[1]{\boldsymbol{#1} }
\newcommand{\ud}{\,\mathrm{d}}

\newtheorem{prop:matrix:approx}{Proposition}
\newtheorem{prop:xtmx}[prop:matrix:approx]{Proposition}
\newtheorem{prop:inv:hilbert}{Lemma}
\newtheorem{prop:crb:alpha}[prop:matrix:approx]{Proposition}
\newtheorem{prop:crb:alpha1}[prop:matrix:approx]{Proposition}
\newtheorem{lemma:varbj}{Lemma}

\begin{document}

\markboth{IEEE TRANSACTIONS ON SIGNAL PROCESSING (TO APPEAR IN Oct. 2012 ISSUE)}{Kar \MakeLowercase{\textit{et al.}}: Cram\'er-Rao Bounds for Polynomial Signal Estimation using Sensors with AR(1) Drift}

\title{Cram\'er-Rao Bounds for Polynomial Signal Estimation using Sensors with AR(1) Drift}

\author{Swarnendu~Kar,~\IEEEmembership{Student~Member,~IEEE,}
        Pramod~K.~Varshney,~\IEEEmembership{Fellow,~IEEE,}
        and~Marimuthu~Palaniswami,~\IEEEmembership{Fellow,~IEEE}
\thanks{S. Kar and P. K. Varshney are with the Department
of Electrical Engineering and Computer Science, Syracuse University, NY 13244 USA, e-mail: \{swkar,varshney\}@syr.edu}
\thanks{M. Palaniswami is with the Department of Electrical and Electronic Engineering, The University of Melbourne, Victoria 3010 Australia, email: palani@unimelb.edu.au}
\thanks{This paper has supplementary downloadable multimedia material available
at http://ieeexplore.ieee.org provided by the authors.}
\thanks{Copyright (c) 2012 IEEE. Personal use of this material is permitted. However, permission to use this material for any other purposes must be obtained from the IEEE by sending a request to pubs-permissions@ieee.org.}}

\maketitle

\begin{abstract}
We seek to characterize the estimation performance of a sensor network where the individual sensors exhibit the phenomenon of drift, i.e., a gradual change of the bias. Though estimation in the presence of random errors has been extensively studied in the literature, the loss of estimation performance due to systematic errors like drift have rarely been looked into. In this paper, we derive closed-form Fisher Information matrix and subsequently Cram\'er-Rao bounds (upto reasonable approximation) for the estimation accuracy of drift-corrupted signals. We assume a polynomial time-series as the representative signal and an autoregressive process model for the drift. When the Markov parameter for drift $\rho<1$, we show that the first-order effect of drift is asymptotically equivalent to scaling the measurement noise by an appropriate factor. For $\rho=1$, i.e., when the drift is non-stationary, we show that the constant part of a signal can only be estimated inconsistently (non-zero asymptotic variance). Practical usage of the results are demonstrated through the analysis of 1) networks with multiple sensors and 2) bandwidth limited networks communicating only quantized observations.

\end{abstract}


%
\IEEEpeerreviewmaketitle

\section{Introduction}
\subsection{The phenomenon of drift}
Sensor networks often consist of a number of sensors deployed for specific tasks such as detection and estimation of the ambient phenomenon \cite{Akyildiz02}. Sensing nodes are equipped with sensing, computation, and inference capabilities.  Sensor networks can potentially deploy a large number of sensors, can measure the environment over time and generate data, from which we seek to extract relevant information regarding the phenomenon. Often while assessing the estimation performance when using such a network, sensor inaccuracies are modeled as independent measurement errors, e.g., i.i.d. Gaussian errors were assumed in \cite{Rib06},\cite{Niu06}. However, it has been widely reported that a variety of sensors used in various applications exhibit systematic errors or drift, e.g., Ground Moving Target Indicator radar sensors \cite{Bar01}, tilt sensors in bridge monitoring applications \cite{Phil08}, \co sensors used for air quality monitoring \cite{Fis07} and salinity sensors for ocean monitoring \cite{Oka05}. For example, gas sensors are known to drift due to other environmental parameters like temperature and humidity \cite{Hol04}.

\subsection{Inference with drifting sensors}
When sensors in a network develop systematic errors with time, inferences based on these observations become increasingly inaccurate. For example, a scheme where each sensor can track their own drift in a collaborative manner is presented in \cite{Mae08} and such a system was shown to accumulate error over time. This necessitates periodic calibration/ registration of the sensors, a procedure that increases the operating cost for the network. Though the issue of drift is often acknowledged in the literature, performance analysis of a network consisting of drifting sensors has not been done before. Our goal in this paper is to characterize the estimation performance of a sensor network in terms of the drift properties of the constituent sensors. 


We consider multiple sensors deployed in a field of interest to monitor a particular environment. 
The sensors observe the same phenomena over $n=1,2,\ldots,N$ time instants and their observations are corrupted by noise and drift. The noisy observations are relayed to a common sink, where the objective is to estimate the parameters governing the observed phenomena. Let $z_{n,m}$ denote the noisy observation of the $m$\sups{th} sensor at the $n$\sups{th} instant and $\bo z_{m}$ denote the vector $[z_{1,m},z_{2,m},\ldots,z_{N,m}]'$. We assume a linear observation model of the form
\begin{align}
\bo z_m =\bo X \bo \beta+\bo \epsilon_m, \quad \bo \epsilon_m\sim\mathcal N(0,\bo \Sigma_m) \label{def:z}
\end{align}
where $\bo x\triangleq \bo X \bo \beta$ and $\bo \epsilon_m$ are $N\times 1$ vectors denoting the  signal and error terms. Here $\bo X$ is assumed to be a $N\times (P+1)$ known matrix (describing the temporal-dynamics) and $\bo \beta$ is the $(P+1)\times 1$ vector denoting an unknown but deterministic signal. The error term $\bo \epsilon_m$ consists of noise and drift components (to be described later) and is assumed to be Gaussian distributed with covariance matrix $\bo \Sigma_m$. The observations $\bo z_m$ are communicated to a sink, whose job is to obtain an accurate estimate of $\beta$. From estimation theory (see, for example, \cite{Kay93} and \cite{Bar01book}), it is well known that the variance of an unbiased estimator is lower bounded by the Cram\'er-Rao Lower Bound (CRB) and that the Maximum-Likelihood (ML) estimator asymptotically (for small errors) attains that bound. In this paper, we derive the closed-form CRB (upto reasonable approximations) for estimating $\bo \beta$ when $\bo \Sigma_m$ corresponds to drift corrupted errors and $\bo X$ corresponds to a polynomial signal. In the subsequent discussion, we provide the motivation and full descriptions of $\bo \Sigma_m$ and $\bo X$.


\subsection{Models of Noise and Drift}
At instant $n$, let $x_n$ and $\epsilon_{n,m}$ denote the common signal magnitude and observation error of sensor $m$. When the effect due to drift is ignored, the measurement noise is often modeled as independent and identically distributed Gaussian noise $w_{n,m}$, (e.g., \cite{Rib06}, \cite{Niu06}), i.e.,
\begin{equation}
z_{n,m}=x_n+w_{n,m}, \quad w_{n,m}\stackrel{\text{i.i.d.}}{\sim}\mathcal{N}(0,\sigma_m^2), \label{model:meas:nodrift}
\end{equation}
where $\sigma_m^2$ is the noise variance. However, in the presence of drift, the observation error has two components: one due to baseline-drift $d_{n,m}$ and the other due to random measurement noise $w_{n,m}$,
\begin{equation}
z_{n,m}=x_n+d_{n,m}+w_{n,m}. \label{model:meas:drift}
\end{equation}
Drift is generally described as a gradual change of the bias of a sensor \cite{Hol04}, \cite{Guo97}. Depending on the specific sensing methodology, various models have been proposed to characterize the drift sequence $\{d_{n,m}\}_{n=1}^{N}$ in frequency and time domains (for a survey, see \cite{Hol04}). Below we describe three commonly used models of drift: 

\emph{1) Frequency domain:} The phase-drift in an oscillator is often modeled in the frequency domain using several powers of frequency (power-law model) \cite{Guo97}, i.e., the power spectral density (PSD) is of the form, $S_m(f)=\sum_{i=-2}^{2} h_{m,i} f^i$, where $h_{m,i}$ are appropriate constants.

\emph{2) Temporal domain - Deterministic: } Drift is also modeled sometimes as a linear movement of the sensor baseline, i.e., $d_{n,m}=a_m+(n-1)b_m$, where the intercept $a_m$ is set to zero after each calibration and slope $b_m$ is assumed to be an unknown constant that is often estimated later and compensated for. Example applications include odor identification using gas-sensor arrays \cite{Pearce98} and air pollution monitoring using gas sensor networks \cite{Tsujita04}.

\emph{3) Temporal domain - Random: } In the broader signal processing literature, auto-regressive moving average (ARMA) processes are often used to describe serially correlated time series, an example of which is drift. As a trade-off between modeling-efficiency and analytical-complexity, drift is often modeled as a first-order autoregressive (AR(1)) process. Example applications include Ground Moving Target Indicator (GMTI) radars (\cite{Bar01}, \cite{Kastella00}), Ring-laser Gyroscopes \cite{Seong00}, Liquid Chromatography \cite{Hayashi94} and sensor networks \cite{Mae08}.

In this paper, we will use the first-order autoregressive model to describe the statistical properties of drift. The AR(1) model characterizes the drift behavior at $m$\sups{th} sensor in terms of an auto-correlation parameter $\rho_m$  (visually, a smaller value of $\rho_m$ means that the baseline drift crosses the zero-line more often and looks more like white noise) and a strength parameter $\sigma_{\delta,m}^2$  (signifying the magnitude of drift),
\begin{align}
\begin{split}
d_{n+1,m}=\rho_m d_{n,m}+\delta_{n,m}, \mbox{ where}\\ 
\rho_m\in[0,1] \mbox{ and } \delta_{n,m}\stackrel{\text{i.i.d.}}{\sim}\mathcal{N}(0,\sigma_{\delta,m}^2). \end{split} \label{model:drift}
\end{align}
Note that, when $\rho=1$, the drift is modeled effectively as a non-stationary random walk, as in \cite{Mae08}.

If we define $\gamma_m$ such that $\sigma_{\delta,m}^2=\gamma_m\sigma_m^2$, our \bq AR(1)+White noise" model for observation error is completely parameterized by $\{\sigma_m^2,\rho_m, \gamma_m\}$. These noise and drift parameters usually have to be identified from the empirical PSD for stationary noise (e.g., \cite{Hayashi94}, \cite{Jakobson2000}) or other time-domain features for non-stationary noise (e.g., \cite{Hazel97}). In this paper, we consider the characterization of the sensing uncertainty in terms of $\{\sigma_m^2,\rho_m,\gamma_m\}$ as part of system identification that must be done prior to an observation cycle. Estimation of these parameters is beyond the scope of this paper. Moreover, within an observation cycle, the drift-parameters are assumed to be constant. If the drift parameters change frequently, our proposed framework must be used in conjunction with periodic system re-identifications.

\subsection{Deterministic signal model}
Often in the sensor network literature, the signal of interest is assumed to be constant over the observation duration, e.g., \cite{Rib06},\cite{Niu06}. This means $x_n=\beta, \forall n$. However, such a signal model may be too simplistic for real applications and we consider a generalization of the form,
\begin{equation}
x_n=\sum_{p=0}^P \beta_p n^p, \qquad \beta_p\in\mathbf{R}, \label{model:sig:poly}
\end{equation}
where $\beta_p$-s are the unknown constants that need to be estimated and $P$ is the order of the polynomial time-series that is assumed to be known. In vector notations, the polynomial signal $\bo x\triangleq [x_1,x_2,\ldots,x_N]'$ assumes a linear form $\bo x =\bo X \bo \beta$, where $\bo X$ is the Vandermonde matrix
\begin{align}
\mathbf X=\begin{bmatrix}
1 & 1 & \cdots & 1 \\
1 & 2 & \cdots & 2^P \\
\vdots & \vdots & \ddots & \vdots \\
1 & N & \cdots & N^P
\end{bmatrix}, \; \mbox{and }
\bo \beta=\left[
\begin{array}{c}
\beta_0 \\
\beta_1 \\
\vdots  \\
\beta_P
\end{array}
\right]. \label{def:XB}
\end{align}

It may be noted that time-varying signals in different applications are approximated by either polynomials or piecewise-polynomials \cite{Unser99splines},\cite{Vett01}. For example, a polynomial regression-based data gathering algorithm for environmental monitoring applications was suggested in \cite{Wang07}. Also, a polynomial spline approximation of stationary random processes was applied to Clarke's model of multipath fading channels in \cite{Zakharov04}.

This completes the description of the signal and observation noise considered in this paper. We intend to derive the closed form CRB (for the estimation of $\beta_p$-s) in terms of the signal ($P,N$) and noise $\{\sigma_m^2,\rho_m,\gamma_m\}$ parameters. This would help us characterize the performance of a sensor network and increase our understanding about its estimation capabilities.

\subsection{Related work}
A related area of work is the study of systematic-bias or model-error estimation schemes using multiple, and sometimes collaborative sensors. In the radar signal processing literature, the process of model-based estimation and subsequent removal of systematic errors prior to target tracking is known as sensor-registration \cite{Dana90}, \cite{Zhou97}. In the weather research literature, the serially-correlated forecasting error arising due to modeling deficiency is often considered separately and tracked alongside the model parameters \cite{Dee00}, \cite{Zupanski06}. In the sensor network literature, drift-aware networks perform learning-based collaborative bias estimation to enhance the effective lifetime of the network \cite{Mae08}, \cite{Takruri11}. However, in this paper, we are focused on the quality of estimation in the presence of systematic errors, rather than techniques on mitigating systematic errors.

Several researchers have studied the Cram\'er-Rao bounds for polynomial (or polyphase) signal estimation in the presence of independent (or correlated) noise. The CRB is usually obtained from the inverse of the Fisher Information Matrix (FIM) \cite{Kay93}. For Additive White Gaussian Noise (AWGN), the large sample approximation of FIM is known to be a multiple of the Hilbert Matrix (e.g., \cite{Li04polych}). The second order approximation was derived in \cite{Peleg91} in the context of polynomial phase signals. For a mixture of additive and multiplicative white Gaussian noise, the large sample FIM was shown in \cite{Swami96mult} to be a scalar multiple of the AWGN case. \emph{Our primary contribution in this paper is the derivation of (approximate) closed-form CRB for polynomial signal estimation in the presence of a mixture of white (measurement noise) and AR(1) Gaussian noise (drift)}. We also discuss the non-stationary case when the autoregressive parameter is equal to $1$. To the knowledge of the authors, polynomial signal estimation in such a mixture of noise has not been considered earlier. As mentioned earlier, the study of estimating polynomial signals in AR(1)+White noise would help us characterize the capability of a sensor network to infer the parameters of a time-varying signal using sensors with drift.

The rest of the paper is organized as follows. In Section \ref{sec:probform}, we formally list the assumptions and set up the problem for two scenarios based on the initial state of the sensor. In Section \ref{sec:mainres}, we consider the single-sensor case and derive large sample approximations of the CRB for $\rho<1$ and $\rho=1$. In Section \ref{sec:multi}, we extend the results to multiple sensors having different drift characteristics. In Section \ref{sec:quant}, we demonstrate the application of the results to a bandwidth limited sensor network that communicates only quantized observations. Concluding remarks and scope of future research are provided in Section \ref{sec:conc}.

\section{Problem Formulation} \label{sec:probform}
To be more specific about our problem framework, we formally state the assumptions here. 
\emph{1) Same phenomenon: } Each sensor is observing the same physical phenomenon (e.g., temperature), which is modeled as a time-polynomial signal. In our framework, if multiple signals are to be sensed (e.g., temperature and humidity), the observations have to be transmitted separately and the parameters independently estimated. 
\emph{2) Known drift statistics: } The drift in each sensor is modeled as AR(1) Gaussian time-series with known statistics - namely the autoregressive and strength parameters, $\rho_m$ and $\gamma_m$ respectively. 
\emph{3) Spatially uncorrelated noise and drift: } We assume that the observations at all the sensors at a particular instant are independent, conditioned on the signal magnitude at that instant. In other words, though the observation noise samples $d_{n,m}+w_{n,m}$ are temporally correlated (due to drift), there is no spatial correlation among them.
\emph{4) Synchronized observations: } The clocks of the sensors are synchronized and they have identical sampling intervals. 
\emph{5) Parallel sensor network with fusion center: } We assume that the sensors do not collaborate among themselves and rather communicate their observations only to the sink. 
\emph{6) Reliable signal transmission: } In Sections \ref{sec:mainres} and \ref{sec:multi}, we assume that the noisy observations are communicated perfectly to the sink without any further distortion. We call this the \emph{full-precision} case which helps obtain a benchmark performance. However, there may be cases when, due to power and bandwidth limitations at the sensor nodes, reliable communication of full-precision observations may be impossible. In such cases, a digital communication based framework in conjunction with efficient channel coding can be used to reliably transmit only a finite number of bits \cite{Rib06}, \cite{Luo05}. In Section \ref{sec:quant}, we will discuss inference using digitized observations where all sensor nodes perform quantization with identical fidelity\footnote{For sensor networks where there is a constraint on system-wide bandwidth (summed across all nodes), different sensors nodes may be assigned different fidelity of quantization based on their observation quality, e.g., \cite{Krasnopeev05}. Though outside the scope of this paper, optimal fidelity assignment for sensors in the presence of drift is a challenging topic worthy of future research.}. With the aforementioned assumptions, our goal in this paper is to derive approximate closed form expressions for the Cram\'er-Rao bounds. 

\subsection{Sensor calibration and noise covariance} \label{sec:noise:cov}
Consider a single sensor with noise and drift properties denoted by $\{\sigma^2, \gamma, \rho\}$ (the subscript $m$ is dropped for most of this subsection). Assume that inference has to be performed from noisy observations $z_n$ at instants $n=1,\ldots,N$. Let $\tau>0$ denote the time-instants elapsed since the sensor was last calibrated, i.e., when the drift component was set to zero by correcting the baseline. Since the duration of inference starts from instant $1$, it follows that the drift sequence was set to zero at instant $1-\tau$ (which is a non-positive index, a slight notational inconvenience). Therefore, the drift sequence proceeds as follows,
\begin{align}
d_{1-\tau}=0,\,  d_{2-\tau}=\delta_1,\, d_{3-\tau}=\rho \delta_1+\delta_2\,\ldots \text{etc.}.
\end{align}
Since each of the drift innovations $\delta_i\sim\mathcal N(0,\gamma\sigma^2)$, we have
\begin{align}
\text{Var}(d_n)&=\gamma\sigma^2(1+\rho^2+\cdots+\rho^{2(\tau+n-1)})=: \gamma\sigma^2 S^\tau_n,
\end{align}
where the summation within the parenthesis is defined as $S^\tau_n$. Though the drift sequence $\{d_n\}_{n=1}^N$ is stationary for large magnitudes of $\tau$,
\begin{align}
\lim_{\tau\rightarrow\infty} S^\tau_n=\frac{1}{1-\rho^2}, \; \rho<1,
\end{align}
it is not stationary in the transient stage, for which,
\begin{align}
S^\tau_1< S^\tau_2<\cdots < S^\tau_N,\quad \tau<\infty.
\end{align}
In this paper, though we would solve the estimation problem for a general $\tau$, we would refer to the limiting cases $\tau=1$ as \emph{calibrated} (C) and $\tau\rightarrow\infty$ as \emph{uncalibrated} (U) respectively.

We define $\bo R$ such that $\gamma\sigma^2 \bo R$ is the covariance matrix of the drift vector $\bo d\triangleq [d_{1},d_{2},\ldots,d_{N}]'$
\begin{align}
\begin{split}
\mathbb{E}[\bo d'\bo d] &= \gamma\sigma^2\begin{bmatrix}
S^\tau_1                  & \rho S^\tau_1              & \rho^2 S^\tau_1  & \cdots  &\rho^{N-1} S^\tau_1\\
\rho S^\tau_1           & S^\tau_2                     & \rho S^\tau_2      & \cdots  & \rho^{N-2} S^\tau_2\\
\rho^2 S^\tau_1       & \rho S^\tau_2              & S^\tau_3             &            & \vdots                      \\
\vdots                       & \vdots                         &                            & \ddots   & \vdots                      \\
\rho^{N-1} S^\tau_1 & \rho^{N-2} S^\tau_2  & \cdots                  & \cdots    & S^\tau_N                
\end{bmatrix}\\
&=:  \gamma\sigma^2\bo R. \label{def:RC}
\end{split}
\end{align}
For the uncalibrated case, the diagonal elements are the same ($S^\tau_n=\frac{1}{1-\rho^2}$), and $\bo R$ takes the shape of the well known stationary matrix  (e.g., \cite{Niu10}),
\begin{align}
\mathbb{E}[\bo d'\bo d] &= \frac{\gamma\sigma^2}{1-\rho^2}\begin{bmatrix}
1 & \rho  & \cdots & \rho^{N-1}\\
\rho & 1 &  \cdots & \rho^{N-2}\\
\vdots & \vdots  & \ddots & \vdots \\
\rho^{N-1} & \rho^{N-2}  & \cdots & 1\\
\end{bmatrix}, \label{def:RU}
\end{align}
sometimes referred to as the Kac--Murdock--Szeg$\ddot{o}$ matrix.

For the $m$\sups{th} sensor, we denote the covariance of the drift sequence by $\bo R_{m}$. Let $\bo \Sigma_{m}$ denote the covariance matrix of the total error (AR(1)+White noise), i.e., $\bo \epsilon_m =\bo w_m + \bo d_m$, so that,
\begin{equation}
\bo \Sigma_{m}=\sigma_m^2\left(\bo I +\gamma_m \bo R_{m}\right), \label{def:Sigma}
\end{equation}
where $\bo I$ is the $N\times N$ identity matrix. 

\subsection{Maximum-Likelihood estimation and Cram\'er-Rao bound}
Given the linear observation model \eqref{def:z} and noise covariance described in Section \ref{sec:noise:cov}, the Maximum-Likelihood (ML) estimator of $\bo \beta$ at the sink is of the form (for a reference on estimation theory, see \cite{Kay93}, \cite{Bar01book}),
\begin{align}
\widehat{\mathbf{\beta}}^{\text{ML}}=\bo J^{-1} \bo X' \sum_{m=1}^M \bo \Sigma_{m}^{-1} \bo z_m, \label{def:ML}
\end{align}
where $\bo J$ is known as the Fisher Information matrix,
\begin{align}
\bo J\triangleq \bo X' \left(\sum_{m=1}^M \bo \Sigma_m^{-1}\right) \bo X,  \label{def:J}
\end{align}

It is well known in estimation theory \cite{Kay93} that within the class of unbiased estimators, the ML estimator of a linear model is optimal in terms of estimation variance \cite{Kay93}. Also, the least possible estimation variance is provided by the Cram\'er-Rao lower bound, $\bo V\triangleq \bo J^{-1}$, so that
\begin{align}
\mathbb E \left[(\beta_p-\widehat \beta_p^{\text{ML}})^2\right] \ge \bo V_{p,p} = \left[\bo J^{-1}\right]_{p,p}, \label{def:CRB}
\end{align}
for $ 0\le p\le P$. Since Equation \eqref{def:CRB} holds with a strict equality for linear models, the CRB is an appropriate performance metric for our problem.

It is unclear from \eqref{def:J} and \eqref{def:CRB} exactly how the estimation variance $\bo V_{p,p}$ depends on the drift parameters $\sigma_m^2,\gamma_m,\rho_m$, signal parameters $P$ and the sample size $N$. \emph{Our goal in this paper is to derive large sample approximations for these Cram\'er-Rao bounds and thereby provide insight into the behavior of estimation performance as it relates to the drift properties of the sensors.}

\section{Main Result: Single Sensor Case} \label{sec:mainres}
Towards obtaining the CRB for the general case of multiple sensors, we first consider the single-sensor scenario and thereby obtain the core results of this paper. Subsequent applications of these core results to the multi-sensor scenario (Section \ref{sec:multi}) and bandwidth limited networks (Section \ref{sec:quant}), as we shall see later, will be somewhat straightforward extensions.

While analyzing the single-sensor scenario, for notational brevity, we drop the sensor-index subscript $m$ from $\sigma_m^2$, $\rho_m$, $\gamma_m$, $\bo R_{m}$ and $\bo \Sigma_{m}$ for the remainder of this section. Note from \eqref{def:J} and \eqref{def:CRB} that, for a single sensor, the Fisher Information matrix is $\bo J=\bo X' \bo \Sigma^{-1} \bo X$ and the Cram\'er-Rao bound is $[\bo V]_{p,p}=[\bo J^{-1}]_{p,p}$. In the following discussion, we first compute the inverses of the disturbance covariance matrices. Next, we use those results to derive the CRB.

\subsection{Inverse of disturbance covariance}
Our goal is to approximate $\bo \Sigma^{-1}$ in a form that are analytically tractable. The exact closed form expression for $\bo \Sigma_U^{-1}$ (for $\tau\rightarrow\infty$) is known to be quite complicated (e.g., p-53 of \cite{Skidmore64}, \cite{Galbraith74}) and we will not use it. The authors have not found a closed form expression for either $\bo \Sigma_C^{-1}$ (the $\tau=1$ case) or $\bo \Sigma^{-1}$ (for general $\tau$) in the literature.

In this paper, we propose an approximation to $\bo \Sigma^{-1}$ that is novel to the best of knowledge of the authors. We suggest the following form for the inverse
\begin{align}
\left( \bo I+\gamma\bo R \right)^{-1} = \bo I -\nu \bo M+\mathcal O(y^N), \label{igRi:app:inM}
\end{align}
where $\nu$ is a constant, $\bo M$ is a matrix with certain structure (to be described later), and $y$ is a quantity less than $1$, so that $\mathcal O(y^N)$ represents a term that vanishes exponentially with sample size $N$. The idea is that, for large $N$, we should be able to use the approximation
\begin{align}
\left( \bo I+\gamma\bo R \right)^{-1} \approx \bo I -\nu \bo M
\end{align}
towards computing the CRB. The specifics of this approximation are laid out in Proposition \ref{pr:matrix:approx}.

\begin{prop:matrix:approx} \label{pr:matrix:approx}
Based on drift parameters $\gamma$ and $\rho$, define the following constants,
\begin{align}
\begin{split}
y&\triangleq \frac{1}{2}\left[\frac{\gamma+1}{\rho}+\rho- \sqrt{\left(\frac{\gamma+1}{\rho}+\rho\right)^2-4}\right], \\
\nu &\triangleq \frac{y\gamma}{\rho (1-y^2)}, \, \mbox{and}\, \kappa\triangleq \frac{y(\rho-y)}{1-\rho y}.
\end{split} \label{def:ynukappa}
\end{align}
Then $\left( \bo I+\gamma\bo R \right)^{-1} =  \bo I-\nu\bo M  + \mathcal{O}(y^N)$, where $\bo M$ has the structure shown in Equation \eqref{def:M:structure}, with
\begin{align}
\begin{split}
a_i&\triangleq 1+y^{2(i-1)}\eta_\tau,  \; b_j\triangleq1+y^{2(j-1)}\kappa, \\
\eta_\tau&\triangleq\frac{1-y^2}{1-\rho y+\varrho_\tau y/\rho}-1,\; \varrho_\tau\triangleq \frac{\rho^{2\tau}}{1+\rho^2+\cdots+\rho^{2\tau-2}}.
\end{split} \label{def:etavarrho}
\end{align}

\begin{figure*}[htb]
\begin{align}
\bo M &=  \left[ \begin{array}{cccccccc}
a_1              & y a_1             & y^2 a_1       & \cdots 	& \cdots &  \cdots                 & y^{N-2}a_1         & y^{N-1} b_1\\
y a_1           & a_2                & y a_2            & \cdots     & \cdots & y^{N-4}a_2         & y^{N-3} b_2        & \vdots          \\
y^2 a_1       & y a_2             & a_3               &        	       &          & y^{N-5} b_3       & \vdots                  & \vdots          \\
\vdots          & \vdots            &                      & \ddots 	& \iddots& \vdots                 & \vdots                  & \vdots          \\
\vdots          & \vdots            &                      & \iddots	& \ddots & \vdots                 &  \vdots                 & \vdots          \\
\vdots          & y^{N-4} a_2  & y^{N-5} b_3  & \cdots 	& \cdots & b_3                     & y b_2                   &  y^2 b_1     \\
y^{N-2}a_1 & y^{N-3} b_2  & \cdots       	   & \cdots 	& \cdots & y b_2                  & b_2                      &  y b_1         \\
y^{N-1}b_1 & \cdots            & \cdots            & \cdots 	& \cdots & y^2 b_1              & y b_1                   & b_1
\end{array} \right], \quad \begin{array}{l}
a_i=1+y^{2(i-1)}\eta_\tau, \\
b_j=1+y^{2(j-1)}\kappa.
\end{array} \label{def:M:structure}
\end{align}
\end{figure*}
\end{prop:matrix:approx}
\begin{IEEEproof} See Appendix \ref{app:prop:1}. \end{IEEEproof}

The constant $y$ can perhaps be more conveniently described as the smaller of the roots of the quadratic equation (both of which are positive) $y^2-\left(\frac{\gamma+1}{\rho}+\rho\right)+1=0$.
It is easy to establish that $y<1$ for $\rho \le 1$ and for all $\gamma$, so that $y^N\rightarrow 0$ for large sample size $N$. Also note that for the two special cases of $\tau=1$ and $\tau=\infty$, $\varrho_C=\rho^2, \eta_C=-y^2$ and $\varrho_U=0, \eta_U=\kappa$.

\emph{Remark \ref{pr:matrix:approx}.1: Identity: } The following identity will be used in several places in this paper and follows from definition \eqref{def:ynukappa} after some algebraic manipulations,
\begin{align}
\gamma y=(\rho-y)(1-\rho y). \label{identity:gy}
\end{align}

\emph{Remark \ref{pr:matrix:approx}.2: } It may be noted that $\bo M$ is defined entirely by the diagonal elements $a_1,a_2,\ldots,a_{\lfloor \frac{N}{2} \rfloor}$ (counted from top) and $b_1,b_2,\ldots,b_{\lceil \frac{N}{2} \rceil}$ (counted from bottom). 
From the expressions for $a_i$ and $b_j$ in Equation \eqref{def:M:structure}, we note that for large $N$, both $a_{\lfloor \frac{N}{2} \rfloor}$ and $b_{\lceil \frac{N}{2} \rceil}$ converge to the same value, namely $1$. That means that it should not matter whether the anti-diagonal elements are expressed in terms of $a_i$-s or $b_j$-s. In Proposition \ref{pr:matrix:approx}, we describe the antidiagonal elements in terms of $b_j$-s just for the sake of being definitive.

\subsection{Computation of Fisher Information Matrix}
We use Proposition \ref{pr:matrix:approx} in \eqref{def:Sigma} and \eqref{def:J} to compute the Fisher Information matrix,
\begin{align}
\bo J &\approx \frac{1}{\sigma^2} \bo X'(\bo I-\nu\bo M) \bo X, \label{J:approx}
\end{align}
which would be inverted later to obtain the CRB-s. We consider the matrices $\bo X'\bo X$ and $\bo X'\bo M\bo X$ individually before summing them up. We would approximate all the elements of these matrices as polynomials in $N$ (i.e., the sample size). This will help derive approximations of the CRB that are correct upto the second order for $\rho \le 1$.

We note from \eqref{def:XB} that $\bo X' \bo X$ is a Hankel (equal skew-diagonal elements) matrix,
\begin{align}
\left[ \bo X' \bo X \right]_{k,l}&=\sum_{n=1}^N n^{k+l}, \quad 0\le k,l\le P. \label{xtx:kl}
\end{align}
It is a well known result, e.g., \cite{Peleg91}, \cite{Swami96mult} that summations of the form \eqref{xtx:kl} can be written as
\begin{align}
\begin{split}
\sum_{n=1}^N n^q =\sum_{i=0}^q \mathcal B_{q,i} N^{q+1-i}, \quad q\ge 0, \quad \mbox{ where} \\
\mathcal B_{q,0}\triangleq \frac{1}{q+1}, \mathcal B_{q,1}\triangleq \frac{1}{2},\mathcal B_{q,2}\triangleq \frac{q}{12}, \mathcal B_{q,3}\triangleq 0,\mbox{etc}. \label{def:Bcal:qi}
\end{split}
\end{align}
A general form for $\mathcal B_{q,i}$ can be found, for example, on p-1, \cite{grad07}. However, all the results in this paper will be established using the first four terms which we have enumerated in \eqref{def:Bcal:qi}.
Similar polynomial expressions can be obtained for $\bo X' \bo M \bo X$.
\begin{prop:xtmx} \label{pr:xtmx} For $0\le k,l\le P$, we have
\begin{align}
\begin{split}
\left[\bo X'  \bo M \bo X \right]_{k,l} &= \sum_{i=0}^{k+l} \mathcal A_{k,l,i} N^{k+l+1-i}+\alpha_{k,l}^{(\tau)}+\mathcal O(N^{k+l+1}y^N),
\end{split} \label{xtmx:kl}
\end{align}
where some leading constants $\mathcal A_{k,l,i}$ and $\alpha_{k,l}^{(\tau)}$ are 
\begin{align}
\begin{split}
\mathcal A_{k,l,0}  &\triangleq  \frac{1+y}{1-y}\frac{1}{k+l+1}, \quad \forall k,l,            \\
\mathcal A_{k,l,1}  & \triangleq \frac{1+2 \kappa-2 y-y^2}{2 (1-y)^2},   \; k+l\ge 1,\\
\alpha_{0,0}^{(\tau)}&\triangleq \frac{-2y+\eta_\tau+\kappa}{(1-y)^2}, \mbox{ etc.} 
\end{split} \label{def:A:alpha}
\end{align}
\end{prop:xtmx}
\begin{IEEEproof} See Appendix \ref{app:xtmx}. \end{IEEEproof}

The exact form of $\mathcal A_{k,l,i}$ and  $\alpha_{k,l}^{(\tau)}$ are provided in Appendix \ref{app:xtmx}.  From  \eqref{xtmx:kl}, it is clear that the Fisher information for calibrated and uncalibrated cases differ only by a constant (dependence on $\tau$ is explicitly indicated). This means that when sample size is large, estimation accuracy will be similar for both the cases. We will elaborate this point later. Next, we use the result in Proposition \ref{pr:xtmx} to derive the CRBs for the cases when $\rho<1$ and $\rho=1$.

\subsection{Cram\'er-Rao Bounds for $\rho<1$}
Using \eqref{xtx:kl},\eqref{def:Bcal:qi} and \eqref{xtmx:kl} in \eqref{J:approx}, we obtain the compact second order expression for the Fisher Information Matrix,
\begin{align}
\bo J &= N \bo E \left[ \xi_0 \bo H+ \frac{\xi_1 \bo e \bo e'+\xi_2^{(\tau)} \bo f \bo f'}{N} +\mathcal O\left(\frac{1}{N^2} \right) \right] \bo E, \label{J:EH}
\end{align}
where $\bo E, \bo H$ are $(P+1)\times(P+1)$ matrices, $\bo e,\bo f$ are $(P+1)$-dimensional vectors and $\xi_0,\xi_1,\xi_2^{(\tau)}$ are constants defined by
\begin{align}
\begin{split}
\bo E &\triangleq \text{diag}\{1,\ldots,N^P\},\; \bo H_{k,l}\triangleq \frac{1}{k+l+1}, \\
\bo e &\triangleq  [1,\ldots,1]',\; \bo f\triangleq  [1,0,\ldots,0]', \\
\xi_0 &\triangleq \frac{1}{\sigma^2}\left[1-\nu\frac{1+y}{1-y}\right], \xi_1\triangleq \frac{1}{\sigma^2}\left[\frac{1}{2}-\nu\mathcal A_1\right],  \\ \xi_2^{(\tau)} &\triangleq -\xi_1-\frac{1}{\sigma^2}\nu \alpha_{0,0}^{(\tau)},
\end{split} \label{def:xiEHef}
\end{align}
where $0\le k,l \le P$. The notation $\mathcal A_1$  is actually $\mathcal A_{k,l,1}$ with the subscripts dropped, since $\mathcal A_{k,l,1}$ does not depend on $k$ and $l$ (see \eqref{def:A:alpha}). $\bo H$ is the well known Hilbert matrix \cite{Peleg91}.

The leading constant $\xi_0$ can be simplified further based on definitions in \eqref{def:ynukappa} and the identity \eqref{identity:gy},
\begin{align}
\xi_0 =\frac{1}{\sigma^2}\left[1+\frac{\gamma}{(1-\rho)^2} \right]^{-1}. \label{xi0}
\end{align}
The CRB will be obtained as the inverse of the FIM described in \eqref{J:EH}. Since $\rho < 1$, we ensure from \eqref{xi0} that $\xi_0 > 0$. Therefore, the FIM of the form \eqref{J:EH} can be inverted assuming $\xi_0 \bo H$ as the dominant term. Such an inversion was performed in \cite{Peleg91} in the context of polynomial-phase signals. We summarize the analysis in \cite{Peleg91} in the form of Lemma \ref{pr:inv:hilbert}.

\begin{prop:inv:hilbert}   \label{pr:inv:hilbert}
\cite{Peleg91}: Let $\bo H$ be the Hilbert matrix and $\bo e,\bo f$ vectors as defined in \eqref{def:xiEHef}. Let $c_0,c_1,c_2$ be constants such that $c_0\neq 0$. Then, for $0\le p\le P$, the diagonal elements are
\begin{align}
\begin{split}
&\left[\left[c_0 \bo H+ \frac{1}{N}\left(c_1 \bo e \bo e'+c_2 \bo f \bo f'\right)\right]^{-1}\right]_{p,p}  = \\
&\quad \frac{K_{P,p}}{c_0 }\left[\frac{1}{2p+1}-\frac{1}{Nc_0}\left(c_1+c_2 L_{P,p}\right)+\mathcal O\left(\frac{1}{N^2} \right) \right],
\end{split} \label{inv:hilbert}
\end{align}
where $K_{P,p}$ and $L_{P,p}$ are defined by
\begin{align}
K_{P,p} &\triangleq  \left[ (P+p+1) \binom{P+p}{p} \binom{P}{p} \right]^2, \; L_{P,p} \triangleq  \left[\frac{P+1}{p+1}\right]^2. \label{def:KL}
\end{align}
\end{prop:inv:hilbert}

The idea behind Lemma \ref{pr:inv:hilbert} is that, for sufficiently large $N$, terms of order $\mathcal O(\frac{1}{N^2})$ in \eqref{inv:hilbert} can be ignored and we obtain approximate closed form expressions for the diagonal elements. The approximation in \eqref{inv:hilbert} is accurate only for small relative magnitude of the second order term, i.e., small magnitudes of $\frac{(2p+1)(c_1+c_2 L_{P,p})}{c_0 N}$. Lemma \ref{pr:inv:hilbert} can be used now to invert \eqref{J:EH}. The Cram\'er-Rao bound (diagonal terms) obtained in such a manner are summarized in Proposition \ref{prop:crb:alpha:lbl}.

\begin{prop:crb:alpha} \label{prop:crb:alpha:lbl}
For $\rho<1$ and sufficiently large $N$, the CRB for estimating $\beta_p$ is
\begin{align}
[\bo V]_{p,p} &\approx \frac{K_{P,p}}{N^{2p+1}\xi_0 } \left[\frac{1}{2p+1}-\frac{1}{N\xi_0} \left(\xi_1+\xi_2^{(\tau)} L_{P,p} \right) \right], \label{crlb:bj}
\end{align}
for $0\le p\le P$.
\end{prop:crb:alpha}

A few remarks due to Proposition \ref{prop:crb:alpha:lbl} are in order.

\emph{Remark \ref{prop:crb:alpha:lbl}.1:  Asymptotic performance: }  As expected, the estimation performance does not depend on transient information ($\tau$) when the number of samples used for inference is sufficiently large. From \eqref{xi0} and \eqref{crlb:bj}, the first order approximation of CRB is
\begin{align}
[\bo V]_{p,p} &\approx \left[1+\frac{\gamma}{(1-\rho)^2} \right] \frac{\sigma^2 K_{P,p}}{N^{2p+1} (2p+1)}, \label{crlb:bjo1}
\end{align}
which is true for any value of $\tau$.

\emph{Remark \ref{prop:crb:alpha:lbl}.2: Equivalent AWGN noise: } For estimation of polynomial signals, the effect of drift is asymptotically (upto first order) equivalent to scaling the measurement noise by a factor of $1+\gamma/(1-\rho)^2$, where $\gamma,\rho$ characterizes the drift properties of a sensor.

\emph{Remark \ref{prop:crb:alpha:lbl}.3: Constant signal: $x_n=\beta_0, \forall n$: }  Simpler and more precise expressions for CRB may be obtained for a constant signal. When $P=0$, we can directly proceed from \eqref{J:EH}. For this case, we have $\bo e \bo e'=\bo f \bo f'=\bo E= \bo H=1$,  hence
\begin{align}
\bo V &\approx \frac{1}{\xi_0 N+\xi_1+\xi_2^{(\tau)}}=\frac{1}{\xi_0 N-\nu \alpha_{0,0}^{(\tau)}/\sigma^{2}}, \label{crlb:const}
\end{align}
where the scalar $\bo V$ denotes the Cram\'er-Rao bound.

\emph{Remark \ref{prop:crb:alpha:lbl}.4: Dependence on $\tau$: } In terms of estimation of a signal, intuitively, the transient state $\tau<\infty$ should be more informative, since the drift starts from a known point, rather than an unknown point. This intuition is corroborated by equation \eqref{crlb:bj}, from which it is easy to derive that $[\bo V]_{p,p}$ increases with $\tau$ (shown below). In conjunction with Remark \ref{prop:crb:alpha:lbl}.1, we can conclude that, though calibration always results in better estimates, the relative gain in performance $([\bo V_U]_{p,p}-[\bo V_C]_{p,p})/[\bo V_U]_{p,p}$) diminishes with sample size. 

\emph{Proof that $[\bo V]_{p,p}$ increases with $\tau$:} Since $\xi_0>0$ (from \eqref{xi0}), it suffices to show that $\xi_2^{(\tau)}$ decreases with $\tau$ (see \eqref{crlb:bj}), or equivalently, $\alpha_{0,0}^{(\tau)}$ increases with $\tau$ (see \eqref{def:xiEHef}, note that $\nu>0$), which is further equivalent to showing that $\eta_\tau$ increases with $\tau$ (see \eqref{def:A:alpha}). The last condition is verified from \eqref{def:etavarrho}, since $\rho_\tau$ is a decreasing function of $\tau$ and $0<\rho,y<1$.

\emph{Remark \ref{prop:crb:alpha:lbl}.5: Approximation region: } The CRB as expressed by \eqref{crlb:bj} is accurate only for cases when the relative magnitudes of the second order terms,
\begin{align}
\epsilon_p=\frac{(2p+1)(\xi_1+\xi_2^{(\tau)} L_{P,p})}{\xi_0 N},\quad 0\le p\le P, \label{app:cond}
\end{align}
are small. This condition helps specify the operating values of $\sigma^2,\gamma,\rho,P$ and $N$ for which the closed-form CRB \eqref{crlb:bj} can be used for performance analysis.

\subsection{Cram\'er-Rao Bounds for $\rho=1$}
For the case when the AR parameter $\rho=1$, the drift phenomenon is a non-stationary even in the limit of $\tau\rightarrow\infty$, since it is a random walk (e.g., \cite{Mae08}) where the bias builds up with time and is unbounded. Hence, the CRB for the uncalibrated scenario is infinity and we consider only the transient case $\tau<\infty$ here. 

For $\rho=1$, the leading term of the Fisher information matrix $\bo J$ \eqref{J:EH} vanishes since \eqref{xi0} implies that $\xi_0=0$. Hence, terms of order $\mathcal{O}(N^{-2})$ need to be considered. In particular, using definitions \eqref{def:ynukappa}, \eqref{def:A:alpha} and \eqref{def:xiEHef}, the constants can further be simplified as
\begin{align}
\begin{split}
\xi_0&=0,\; \xi_1=0, \; \varrho_\tau=\frac{1}{\tau},\; \widetilde\gamma\triangleq\sqrt{1+4\gamma^{-1}}, \\
\xi_2^{(\tau)}&=\frac{1}{\sigma^2}\left[\frac{2}{\widetilde\gamma+1} \left(1+\frac{2\tau}{\widetilde\gamma-1}\right)\right]^{-1},
\end{split} \label{const:simple:r1}
\end{align}
and we subsequently note (see Appendix \ref{app:fim:crb:r1}) that the FIM is of the structure depicted in equation \eqref{JC:alpha1},
\begin{align}
\begin{split}
\bo J & = \bo E \begin{bmatrix}
\xi_2^{(\tau)} &  \frac{1}{ N}\xi_3^{(\tau)}\bo f_P'+\mathcal O\left(\frac{1}{N^2} \right)           \\
*     & \frac{1}{N}\bo D \left[\xi_4 \bo H_P+\frac{1}{N}\bo F+\mathcal O\left(\frac{1}{N^2}\right)\right] \bo D
\end{bmatrix} \bo E,  \\
&\qquad\qquad \mbox{ with } \bo F=\xi_5 \bo e_P \bo e_P'+\xi_6^{(\tau)} \bo f_P \bo f_P',
\end{split} \label{JC:alpha1}
\end{align}
with the matrix $\bo D$ and constants $\xi_3^{(\tau)}, \xi_4, \xi_5, \xi_6^{(\tau)}$ defined as
\begin{align}
\begin{split}
\bo D& \triangleq \text{diag}\{ 1,\ldots,P \},\; \xi_3^{(\tau)}\triangleq -\frac{1}{\sigma^2}\nu \alpha_{1,0}^{(\tau)},\; \xi_4\triangleq \frac{1}{\sigma^2\gamma}, \\
\xi_5&\triangleq -\frac{1}{\sigma^2} \frac{y^2}{(1-y)^3},\; \xi_6^{(\tau)}\triangleq -\xi_5-\frac{1}{\sigma^2}\nu \alpha_{1,1}^{(\tau)},
\end{split} \label{def:xi34}
\end{align}
and $\bo e_P,\bo f_P$ and $\bo H_P$ being the $P$-dimensional equivalents of $\bo e$, $\bo f$ and $\bo H$ (defined in \eqref{def:xiEHef}) respectively.
The CRB is obtained from the FIM \eqref{JC:alpha1} by using block-inversion and Lemma \ref{pr:inv:hilbert} (see Appendix \ref{app:fim:crb:r1}). 

\begin{prop:crb:alpha1} \label{prop:crb:alpha1:lbl}
For $\rho=1$ and sufficiently large $N$, the CRB for estimating $\beta_p$, $[\bo V]_{p,p}$ is equal to
\begin{align}
\begin{split}
&[\bo V]_{p,p}\approx\left\{ \begin{array}{rl}
 \frac{1}{\xi_2^{(\tau)}}\left[1+\frac{1}{N} \frac{P^2\left(\xi_3^{(\tau)}\right)^2}{\xi_2^{(\tau)} \xi_4} \right], & p=0,  \\
 \frac{ K_{P-1,p-1}}{N^{2p-1}p^2 \xi_4}
\left[ \frac{1}{2p-1} -\frac{1}{N} \widetilde \xi \right], &1\le p\le P,
\end{array} \right. \\
&\qquad \mbox{with }\widetilde \xi = \frac{1}{\xi_4}\left(\xi_5+ \left(\xi_6^{(\tau)}-\frac{\left(\xi_3^{(\tau)}\right)^2}{\xi_2^{(\tau)}}\right)\frac{P^2}{p^2}\right).
\end{split} \label{crlb:bj:alpha1}
\end{align}
\end{prop:crb:alpha1}

A few remarks due to Proposition \ref{prop:crb:alpha1:lbl} are noted below.

\emph{Remark \ref{prop:crb:alpha1:lbl}.1: Constant Signal: $x_n=\beta_0, \forall n$: } For a random walk drift, a constant signal can only be estimated inconsistently with asymptotic variance
\begin{align}
\bo V & \approx \frac{1}{\xi_2^{(\tau)}}=\frac{2 \sigma^2}{\widetilde\gamma+1} \left(1+\frac{2\tau}{\widetilde\gamma-1}\right). \label{crlb:const:alpha1}
\end{align}
It is explicitly seen from \eqref{crlb:const:alpha1} that the variance increases with white noise variance $\sigma^2$, the drift strength $\gamma$ (note that $\widetilde \gamma$ decreases with $\gamma$, see \eqref{const:simple:r1}), and the time since last calibration, $\tau$. This phenomenon of inconsistent estimation can be intuitively understood as follows. Even without the white noise component in \eqref{model:meas:drift}, the observation never captures independent readings of $\beta_0$. Rather, the samples are
\begin{align}
\begin{split}
z_1&=\beta_0+d_\tau+w_1, \\
z_2&=\beta_0+d_{\tau+1}+w_2=(\beta_0+d_\tau)+\delta_{\tau+1}+w_2, \mbox{ etc.},
\end{split}
\end{align}
which means that $\beta_0+d_\tau$ appears together in all subsequent observations, thereby making $\beta_0$ indistinguishable from $d_\tau$. Since $d_\tau$ has a finite variance, $\beta_0$ can only be estimated inconsistently. 

\emph{Remark \ref{prop:crb:alpha1:lbl}.2: Time Features: } For other parameters whose effect on the signal vary with time, i.e., $\beta_p \mbox{ for } p\ge 1$, the CRB is (to first order) equivalent to the case of estimating the derivative of the signal in drift-only noise. We note that the derivative of the signal is $x'(t)=\sum_{p=1}^{P} p\beta_p t^{p-1}$ and the forward difference (see \eqref{model:meas:drift}),
\begin{align}
z_{n+1}-z_{n}&=x_{n+1}-x_{n}+d_{n+1}-d_{n}+w_{n+1}-w_{n} \nonumber \\
&=x'(n) (\Delta T)+\delta_n+w_{n+1}-w_{n} ,
\end{align}
which is an estimator of the derivative, contains independent drift innovations $\delta_n$ with variance $\gamma\sigma^2$. Hence in Proposition \ref{prop:crb:alpha1:lbl}, the variance of estimating $\beta_p$ is scaled down by a factor of $p^2$ compared to the equivalent case with white noise $\gamma \sigma^2$.

\emph{Remark \ref{prop:crb:alpha1:lbl}.3: Approximation region: } The CRB as expressed by \eqref{crlb:bj:alpha1} is accurate only for cases when the relative magnitude of the second order terms,
\begin{align}
\begin{split}
\epsilon_p=\left\{ \begin{array}{rl}
 \frac{1}{N }\frac{P^2\left(\xi_3^{(\tau)}\right)^2}{\xi_2^{(\tau)} \xi_4}, & p=0,  \\
 \frac{2p-1}{N }\widetilde \xi, &1\le p\le P.
\end{array} \right.
\end{split} \label{app:cond:r1}
\end{align}
are small. This condition helps specify the operating values of $\sigma^2,\gamma,P$ and $N$ for which the closed-form CRB \eqref{crlb:bj:alpha1} can be used for performance analysis.

\subsection{Numerical Results - Maximum Relative Error} \label{sec:disc:num}
In the remainder of this section, we demonstrate the accuracy of the closed-form performance approximations using representative examples. We shall consider the approximations expressed in both the forms of Fisher Infomation matrix  (Equations  \eqref{J:EH} and \eqref{JC:alpha1}, with $\mathcal{O}\left(\frac{1}{N^2}\right)$ terms truncated) and Cram\'er-Rao bounds (Equations \eqref{crlb:bj} and  \eqref{crlb:bj:alpha1}). 

We use the metric called Maximum Relative Error (MRE) which was also used in \cite{Peleg91}. This involves computing both the exact \eqref{def:CRB} and approximate CRB-s. The relative deviations of the approximations are then calculated for all $0\leq p\leq P$ and the largest one is called the MRE. We denote
\emph{1) } $[\bo V^\text{TH}]_{p,p}$ as the theoretical CRB as in \eqref{def:CRB} and 
\emph{2) } $[\bo V^\text{AP}]_{p,p}$ as the approximate CRB derived either through 
\emph{2a) FIM approx:} inversion of the intermediate FIM, i.e., $[\bo V^\text{AP}]_{p,p}=[\bo J^{-1}]_{p,p}$, in Equations \eqref{J:EH} and \eqref{JC:alpha1} with $\mathcal{O}\left(\frac{1}{N^2}\right)$ terms truncated, or
\emph{2b) CRB approx:} final Cram\'er-Rao bounds in Equations \eqref{crlb:bj} and  \eqref{crlb:bj:alpha1}.
Then the maximum relative error is defined by
\begin{align}
\mbox{MRE}&=\max_{p=0,1,\cdots,P} \frac{| [\bo V^\text{TH}]_{p,p}-[\bo V^\text{AP}]_{p,p} | }{[\bo V^\text{TH}]_{p,p}}.
\end{align}
In short, the MRE summarizes the approximation error over all components of the parameter vector. 

\begin{figure}
\centering
\subfigure[Stationary drift ($\rho<1$) and calibrated sensors.]{
    \includegraphics[width=\figsza \columnwidth]{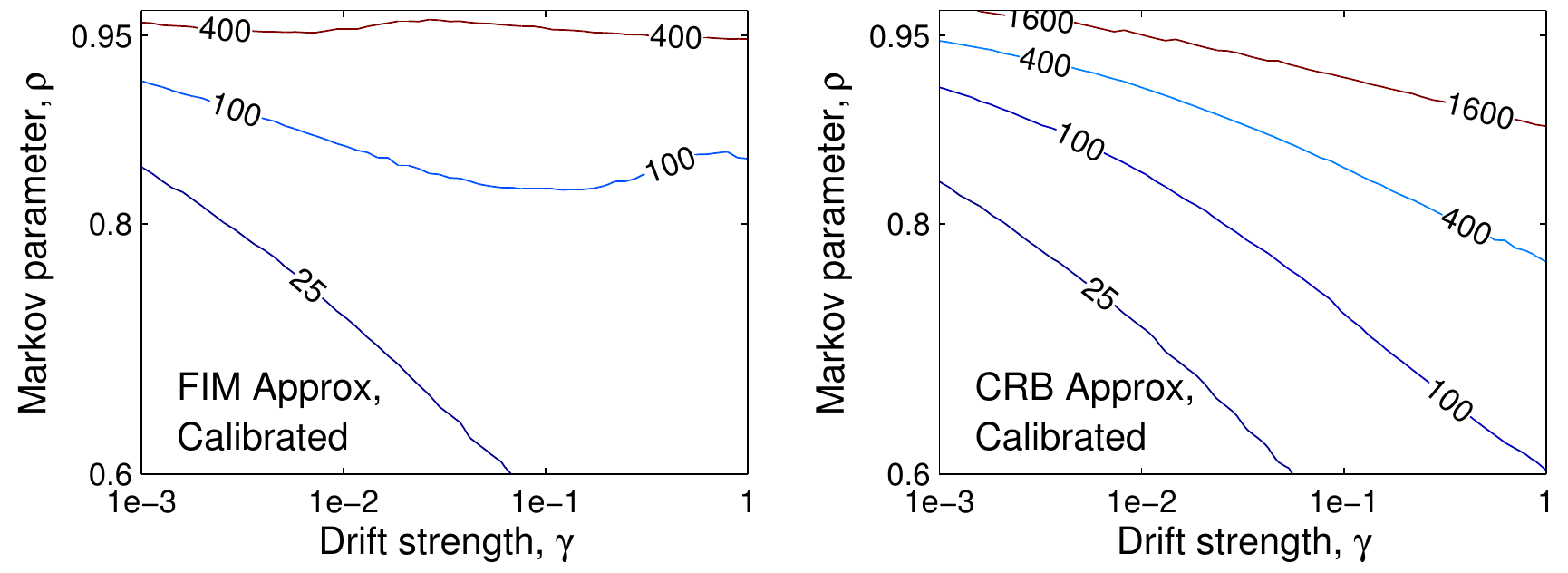}
\label{fig:mre:s0}
}
\subfigure[Stationary drift ($\rho<1$) and uncalibrated sensors.]{
    \includegraphics[width=\figsza \columnwidth]{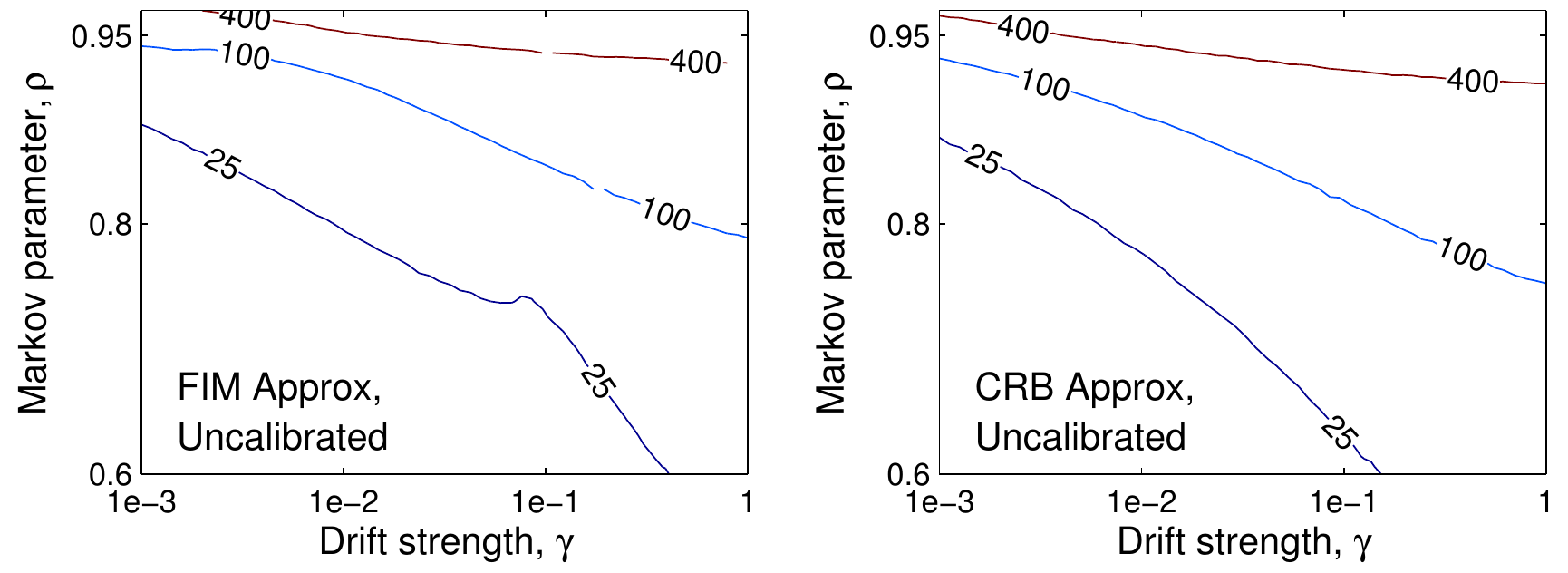}
\label{fig:mre:s1}
}
\subfigure[Non-stationary drift ($\rho=1$).]{
    \includegraphics[width=\figszb \columnwidth]{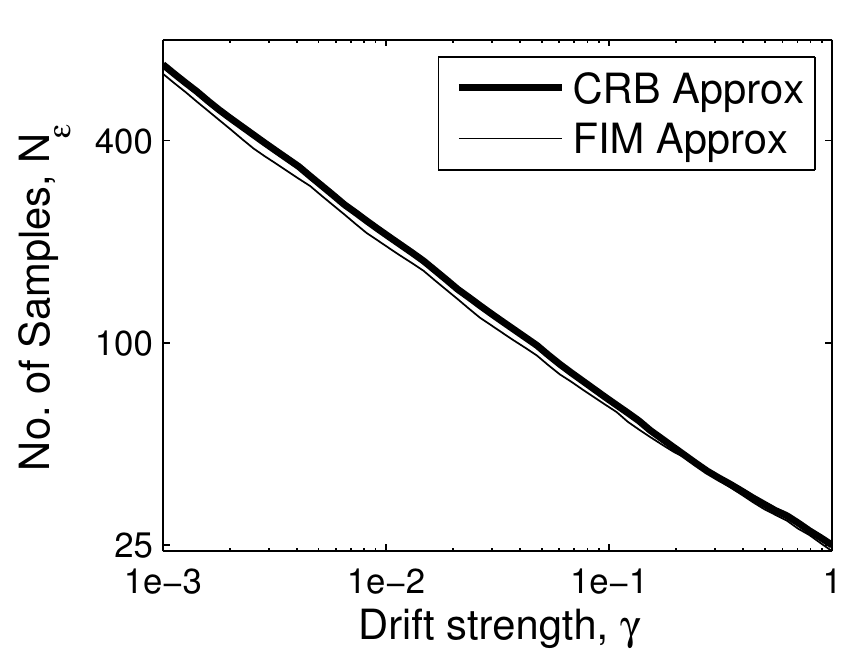}
\label{fig:mre:r1}
}
\caption{Sample size $N_\epsilon$ required for 95\% accurate performance prediction}
\label{fig:mre}
\end{figure}

As an example, we consider the estimation of a cubic-polynomial signal, i.e., $P=2$. Since our performance bounds are asymptotically accurate, it is expected that the MRE will decrease with increasing sample size $N$. In Figure \ref{fig:mre}, we display the sample size $N_\epsilon$ required for 95\% accuracy, or in other words, $\mbox{MRE}<\epsilon=0.05$. Since MRE is a ratio of variances, it does not depend on the measurement noise variance $\sigma^2$. While considering the  $\rho<1$ case, we have displayed the wide parameter region $\gamma\in[10^{-3},1]$, $\rho\in[0.6,0.97]$ in Figures \ref{fig:mre:s0} and \ref{fig:mre:s1} - which demonstrate the uncalibrated ($\tau\rightarrow\infty$) and calibrated ($\tau=1$) scenarios respectively. The  $\rho=1, \tau=1$ case is demonstrated in \ref{fig:mre:r1}, where we have displayed the parameter region $\gamma\in[10^{-3},1]$.

Figure \ref{fig:mre} provides useful guidance on the applicability and limitations of the performance bounds derived in this paper. Firstly, the performance bounds are found to be reasonably accurate over a wide range of possible parameter values and for a moderate number (tens and hundreds) of samples. For accurate performance prediction in stationary drift (Figures \ref{fig:mre:s0} and \ref{fig:mre:s1}), higher values of drift-autocorrelation ($\rho$) and drift-strength ($\gamma$) generally requires larger observation durations ($N_\epsilon$). This is predicted by the approximation-region condition in \eqref{app:cond}, since the denominator $\xi_0$  (see \eqref{xi0}) is inversely proportional to $\rho$ and $\gamma$. For the random-walk drift scenario (Figure \ref{fig:mre:r1}), higher value of drift-strength $\gamma$ requires smaller observation durations ($N_\epsilon$) for accurate performance predictions. This too, can be explained from approximation-region condition in \eqref{app:cond:r1}, where, with the aid of definitions \eqref{const:simple:r1} and \eqref{def:xi34}, it can be established that $\epsilon_p=\mathcal O(\gamma^{-1/2}) $ for small $\gamma$. This also partially explains the log-linear relation between $N_\epsilon$ and $\gamma$ in Figure \ref{fig:mre:r1}.

We have so far only considered the estimation performance of a single sensor that is able to reliably communicate its observations to the sink without any distortion. In Sections \ref{sec:multi} and \ref{sec:quant}, we discuss extensions to the multiple sensor framework and bandwidth limited networks.

\section{ Multiple Sensors } \label{sec:multi}
In this section, we consider the application of the results in Section \ref{sec:mainres} to multiple sensors with different noise and drift parameters $\{\sigma_m^2,\gamma_m,\rho_m\}$. Since the sensor noise and drift are independent across the sensors (Assumption 3), the Fisher Information for $\bo \beta$ is equal to the sum of individual FIM-s (see equation \eqref{def:J}), $\bo J=\bo J_1+\bo J_2+\cdots+\bo J_M$.
Hence the expressions for FIM (given by \eqref{J:EH} and \eqref{JC:alpha1}) and subsequently CRB  (given by \eqref{crlb:bj} and \eqref{crlb:bj:alpha1}) can be extended with the following change in definitions,
\begin{align}
\xi_{\#,\text{eff}}\triangleq \sum_{m=1}^M \xi_\#(\sigma_m^2,\gamma_m,\rho_m), \mbox{ for } [\#]=0,1,\ldots,6, \label{xi:eff}
\end{align}
where $\xi_\#(\sigma^2,\gamma,\rho)$ can be thought of as a function of its arguments as defined in \eqref{def:xiEHef} and \eqref{def:xi34} and $\xi_{\#,\text{eff}}$ denote the effective value of the constant.

As an illustration of this extension, we consider an example where noise and drift parameters are uniformly (randomly) distributed over a given range, say,
\begin{align}
\rho_m \in [\rho_l,\rho_u],\; \sigma_m^2 \in [\sigma_l^2,\sigma_u^2] \mbox{ and } \gamma_m \in [\gamma_l,\gamma_u]. \label{sec:multi:range}
\end{align}
When the number of sensors $M$ is large, \eqref{xi:eff} can be further approximated by substituting the summations by integrals,
\begin{align}
\xi_{\#,\text{eff}}&=M \mathbb E \left[ \xi_\#\right] \nonumber \\
&=\frac{M}{\Delta \rho \Delta \sigma^2 \Delta \gamma}  \int_{\sigma_l^2}^{\sigma_u^2} \int_{\gamma_l}^{\gamma_u} \int_{\rho_l}^{\rho_u} \xi_\# \ud \rho \ud \gamma \ud \sigma^2. \label{xi:eff:int}
\end{align}
We would refer to the CRB derived using the constants $\xi_{\#,\text{eff}}$ (arising out of integration) as the \emph{Average-CRB}. We expect the Average-CRB to be an effective indicator of system performance for large number of sensors $M$, as we demonstrate with some numerical results below.

\subsection{Numerical Results}
\begin{figure}
\begin{center}
    \includegraphics[width=\figsza \columnwidth]{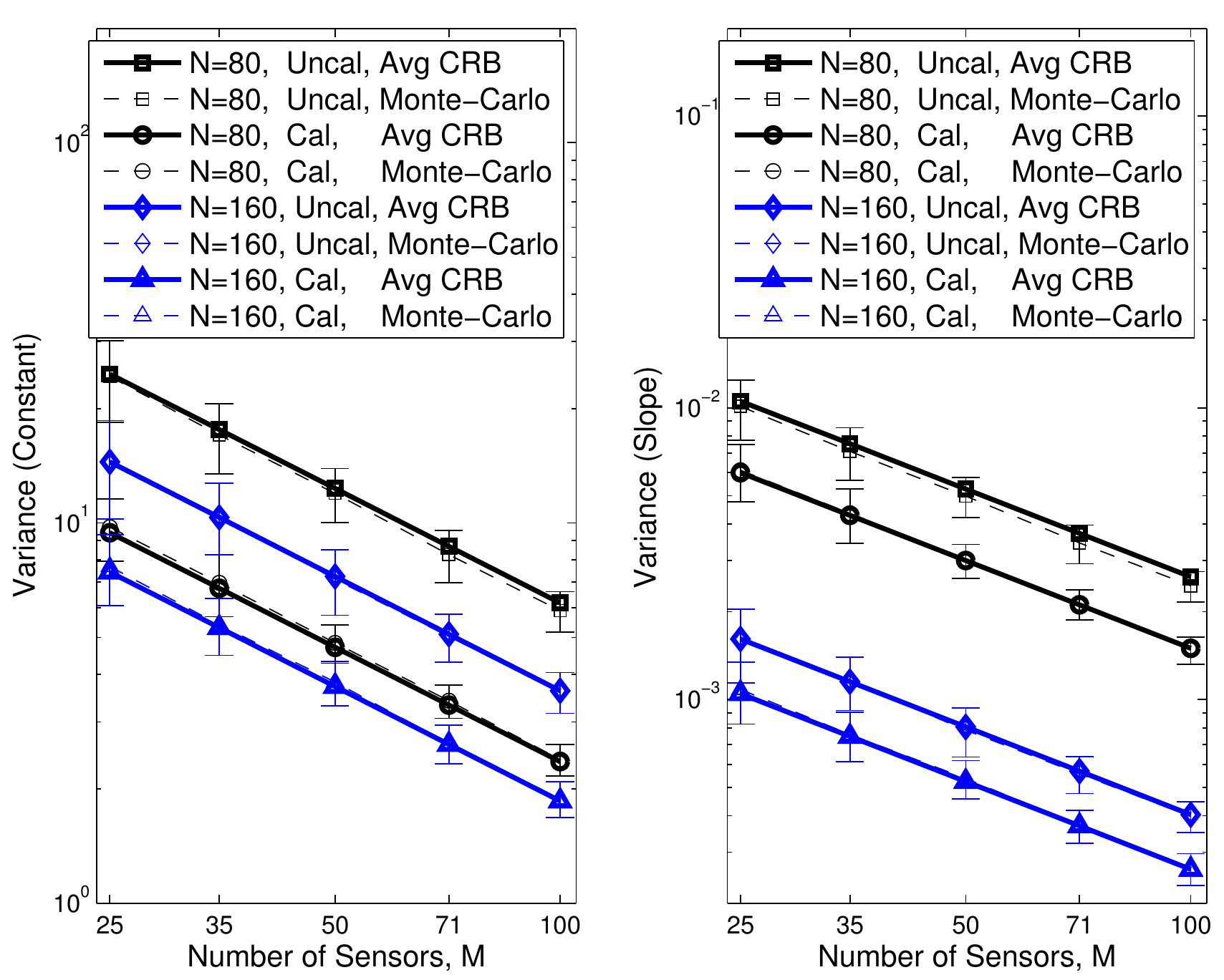}
  \caption{Average performance for multi-sensor systems for $\rho\in[0.85,0.95]$.}
  \label{fig:mc:multi:lin}
  \end{center}
\end{figure}

\begin{figure}
\begin{center}
    \includegraphics[width=\figsza \columnwidth]{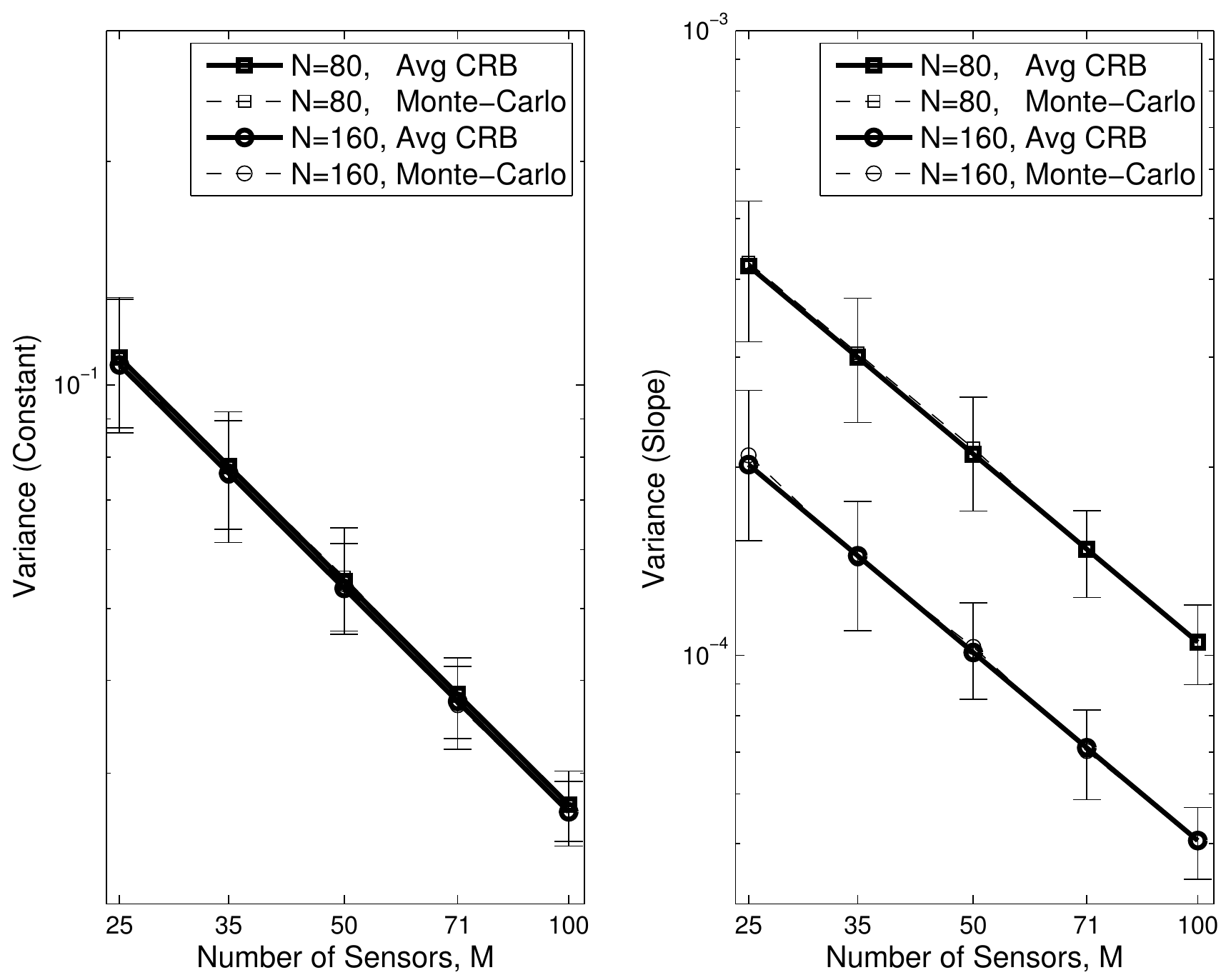}
  \caption{Average performance for multi-sensor systems for $\rho=1$.}
  \label{fig:mc:multi:lin:r1}
  \end{center}
\end{figure}

We demonstrate the results for both $\rho<1$ and $\rho=1$ cases. The simulation setup is described next.
\emph{1) } We consider a linear signal, i.e., $P=1$. The parameters to be estimated are the constant term $\beta_0$ and the slope term $\beta_1$.
\emph{2) } We consider multi-sensor systems with the number of sensors, $M$, starting from $25$ and going upto $100$.
\emph{3) } For the $\rho<1$ scenario, for each sensor, the drift parameters are randomly selected by choosing the parameters from the range,
\begin{align}
\rho_m \in [.85,.95],\; \sigma_m^2 \in [72,288] \mbox{ and } \gamma_m \in [.6,2.4], \label{sec:mc:par:range}
\end{align}
which is close to an estimated spectrum in \cite{Hayashi94}.
Within the $\rho<1$ scenario, we simulate both calibrated ($\tau_m=1,\, \forall m$) and uncalibrated ($\tau_m=\infty,\, \forall m$) cases. In one simulation, we assume that all $M$ sensors are calibrated while on another simulation we assume that none of them are calibrated.
\emph{4) } For the $\rho=1$ scenario, the drift parameters are randomly selected by choosing the parameters from the range,
\begin{align}
\sigma_m^2 \in [2,12] \mbox{ and } \gamma_m \in [.05,.3]. \label{sec:mc:par:range:r1}
\end{align}
\emph{5) } For each realization of an $M$-sensor network, ML-estimation of the constant and slope parameters were performed and the error variances were averaged over $10^5$ Monte-Carlo trials (realizing the measurement noise and sensor drift).
\emph{6) } Several ($10^3$) Monte-carlo trials (realizing $M$-sensor networks with $\sigma^2,\rho_m,\gamma_m$ chosen from above parameter range) are performed and the average and 95\% confidence interval of the error variances are observed. 
\emph{7) } We repeat the experiments for sample sizes $N=80$ and $160$. These samples sizes were chosen to ensure moderate computational (Monte-Carlo) effort.
Since we consider a small number of samples, we have used the expressions for Fisher Information Matrix (Equations \eqref{J:EH} and \eqref{JC:alpha1}) to predict the estimation error variance of the constant $\beta_0$ and slope $\beta_1$ portion of a signal. The results are displayed in Figures \ref{fig:mc:multi:lin} and \ref{fig:mc:multi:lin:r1}, of which we make some comments below.

Firstly, an approximation to the system performance using \eqref{xi:eff:int} is seen to be fairly accurate, as depicted in Figures \ref{fig:mc:multi:lin} and \ref{fig:mc:multi:lin:r1}. In all cases, the variance is inversely proportional to the number of sensors, as depicted by the log-linearity of all the curves. The (average) performance prediction improves (error-bars becomes shorter) for higher sample sizes, since the summation \eqref{xi:eff} is better approximated by the integral \eqref{xi:eff:int} for large $M$. This means that just by knowing the range of the parameter values and the number of sensors, we can have a good understanding about the estimation performance of the entire sensor network. Secondly, from Figure \ref{fig:mc:multi:lin}, we note the intuitive phenomenon that the performance using calibrated sensors is better than that using uncalibrated sensors. Also, though for $N=80$, the performance gap is quite large, the gap between calibrated and uncalibrated cases narrows down for $N=160$. This corroborates Remark \ref{prop:crb:alpha:lbl}.4, where we noted that the relative performance gain diminishes for higher sample sizes. For the  $\rho=1$ case, from Figure \ref{fig:mc:multi:lin:r1}, we note that increasing the sample size does not help in estimation of the constant portion of the signal. This is due the inconsistency property of the non-stationary drift model, as described in Remark \ref{prop:crb:alpha1:lbl}.1.


\section{Quantized observations } \label{sec:quant}
\begin{figure}
\begin{center}
    \includegraphics[width=\figszc \columnwidth]{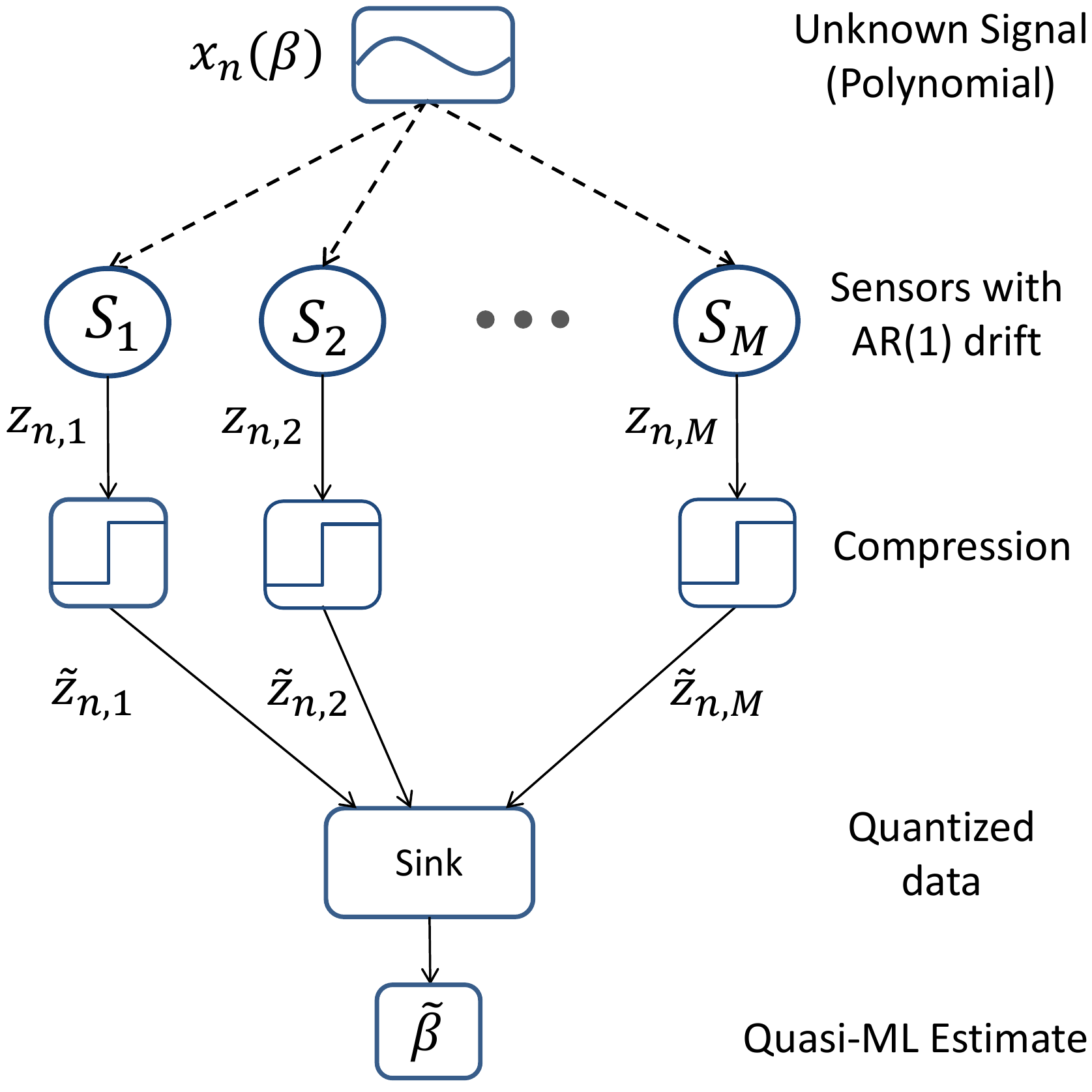}
  \caption{Bandwidth constrained sensor network where inference is performed using quantized data.}
  \label{fig:quantized}
  \end{center}
\end{figure}

As another application of the results in Section \ref{sec:mainres}, we consider the performance characterization of a sensor network where resource (bandwidth, power, etc.) available for communication is limited \cite{Rib06}, \cite{Luo05}. Bandwidth constraints preclude the reliable communication of full-precision observations from the individual sensor nodes to the sink. In such situations, the observations must be compressed and digitized prior to transmission \cite{Gold05}. In this section, we digitize each observation by using uniform quantization, as will be discussed shortly. Since the reliability of the transmission of digitized observations may be further enhanced through the use of efficient channel coding (error-control codes), it is reasonable to assume that once the quantization is performed, there is no further deterioration of the observation-quality \cite{Rib06}. In other words, the quantized observations are assumed to be available error-free at the sink. The schematic diagram of such a system is depicted in Figure \ref{fig:quantized}. \emph{Our goal in this section is to predict the estimation performance of the sensor network composed of sensors with drift, when only quantized observations are available for inference at the sink.}

We use uniform quantization \cite{Krasnopeev05}, in which each of the observations $z_{n,m}$ are quantized uniformly with $l$-bits. Assuming that the observations are bounded $z_{n,m}\in[U_0,U_1]$ and the quantization thresholds ${a_j}\in[U_0,U_1]$, $j=0,\ldots,2^l$ are uniformly spaced such that $a_{j+1}-a_j=(U_1-U_0)/(2^l-1)\triangleq \Delta$, the expected distortion due to the quantization process is
\begin{align}
\sigma_Q^2\triangleq \mathbb E[(\widetilde{z}_{n,m}-z_{n,m})^2]=\frac{\Delta^2}{12}, \quad \forall n,m.
\end{align}
Since the quantization noise $\widetilde{z}_{n,m}-z_{n,m}$, in general, is neither independent across time nor Gaussian distributed, the Maximum-Likelihood estimator is difficult to design. However, the quasi-ML estimator is easier to implement, which is designed on the assumption that the quantization noise is i.i.d. Gaussian with variance $\sigma_Q^2$. We will use the quasi-ML estimator, described below, to perform inference using quantized observations.

Effectively, the quasi-ML estimator assumes that the noise at individual sensor nodes has an added component $\sigma_Q^2\bo I$, with the total covariance being (compare with \eqref{def:Sigma})
\begin{align}
\widetilde{\mathbf\Sigma}_{*,m}&\triangleq\sigma_Q^2\bo I+\mathbf\Sigma_{*,m} \nonumber \\
&=\sigma_Q^2\bo I+\sigma_m^2 (\mathbf I +\gamma_m\mathbf R _{*,m} ) \nonumber \\
&\triangleq\widetilde{\sigma}_m^2\left(\mathbf I +\widetilde{\gamma}_m\mathbf R _{*,m}\right), \label{def:Sigma:quant}
\end{align}
where $\widetilde{\sigma}_m,\widetilde{\gamma}_m$ satisfy 
$\widetilde{\sigma}_m^2\triangleq \sigma_m^2+\sigma_Q^2$ and  $\widetilde{\gamma}_m\triangleq\frac{\gamma_m}{1+\sigma_Q^2/\sigma_m^2}$.
Accordingly, the quasi-ML estimator is (compare to \eqref{def:ML}),
\begin{align}
\begin{split}
\widetilde{\bo \beta}^{\text{ML}}&=\widetilde{J}^{-1} \bo X' \sum_{m=1}^M \widetilde{\bo \Sigma}_{m}^{-1} \widetilde{\bo z}_m, \mbox{where} \\
\widetilde J &\triangleq \bo X^T \left(\sum_{m=1}^M \widetilde{\bo \Sigma}_m^{-1}\right) \bo X.
\end{split}
\end{align}
Note that, in the limiting case when quantization errors are small (equivalently, large $l$), we have $\widetilde{\sigma}_m^2\rightarrow \sigma_m^2,\widetilde{\gamma}_m\rightarrow \gamma_m$ and the quasi-ML-estimator is identical to the ML-estimator. 

We would refer to the CRB derived using modified noise parameters $\{\widetilde{\sigma}_m^2,\widetilde{\gamma}_m,\rho_m\}$ as the \emph{Modified-CRB}. We expect the Modified-CRB to be an effective indicator of system performance for moderate to large number of quantization levels $2^l$. Similar to Section \ref{sec:multi}, with certain change in definitions, the expressions for FIM (given by \eqref{J:EH} and \eqref{JC:alpha1}) and subsequently CRB  (given by \eqref{crlb:bj} and \eqref{crlb:bj:alpha1}) can be extended to obtain the Modified-CRB,
$\xi_{\#,\text{eff}}\triangleq \sum_{m=1}^M \xi_\#(\rho_m,\widetilde\sigma_m^2,\widetilde\gamma_m) $,
where $\xi_\#(\rho,\sigma^2,\gamma)$ can be thought of as a function of its arguments as defined in \eqref{def:xiEHef} and \eqref{def:xi34} for $[\#]=0,1,\ldots,6$. We show some numerical results below to corroborate the effectiveness of Modified-CRB as an efficient performance predictor.

\subsection{Numerical Results}
The simulation setup is described as follows. 
\emph{1) } A sample size of $N=400$ and network size of $M=25$ nodes was considered. 
\emph{2) } The noise and drift parameters of the sensor nodes are chosen by uniformly spacing them in the range,
\begin{align*}
72&=\sigma^2_1 <\sigma^2_2<\cdots<\sigma^2_M=288, \\
0.6&=\gamma_1 <\gamma_2<\cdots<\gamma_M=2.4, \mbox{and} \\
0.85&=\rho_1 <\rho_2<\cdots<\rho_M=0.95.
\end{align*}
Both calibrated ($\tau_m=1,\, \forall m$) and uncalibrated ($\tau_m=\infty,\, \forall m$) cases were considered. 
\emph{3) } We assume a linear signal, i.e., $P=1$ with the constant term $\beta_0=400$ and the slope term $\beta_1=0.9$. 
The range of the observations to be quantized was chosen to be $U_0=0, U_1=1200$, beyond which the observations were clipped. The range was deliberately chosen large so that clipping (which is another source of distortion which we have not modeled) does not occur frequently.
\emph{4) } Uniform quantization is performed using $l=5,6,\ldots,9$ bits per observation.
\emph{5) } Quasi-ML estimation of the constant and slope parameters was performed and the error variances were averaged over $10^8$ Monte-Carlo trials (realizing the measurement noise and sensor drift). The 95\% confidence interval of the error variances were also observed. 

\begin{figure}
\begin{center}
    \includegraphics[width=\figsza \columnwidth]{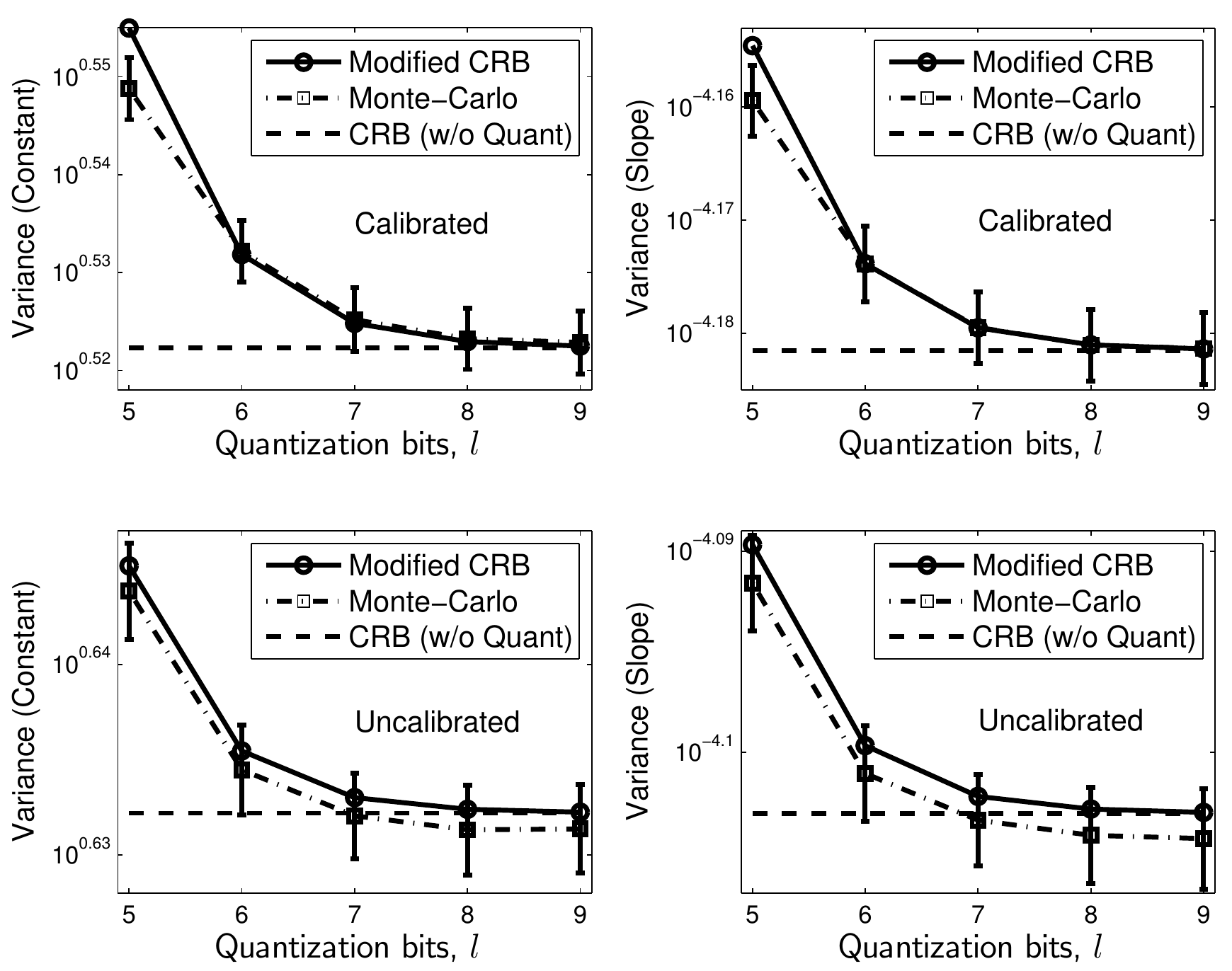}
  \caption{Cram\'er-Rao bounds with quantized observations.}
  \label{fig:mc:quant}
  \end{center}
\end{figure}

The results of Monte-Carlo simulations are compared to theoretical predictions from Modified-CRB and displayed in Figure \ref{fig:mc:quant}. The full-precision CRB is also displayed in the figure (labeled w/o Quant), marking the convergence of Modified-CRB in the large-$l$ regime. The actual estimation variance of both the constant and slope parameters (of the linear signal) seem to agree, with reasonable accuracy, to the Modified-CRB.

\section{Conclusion} \label{sec:conc}
In this paper, we have derived approximate bounds for the estimation accuracy of polynomial signals using sensors that exhibit drift in addition to having measurement errors. This is important since sensor drift, or loss of calibration with time, is a major problem in many applications. The theoretical closed form expressions are validated through numerical results. As future work, we intend to derive  performance measures for other signal models, e.g., stochastic models rather than deterministic. Additionally, it may be interesting to investigate the impact of spatial correlation (i.e, relaxing Assumption 3) on estimation performance. More bit-efficient quantization schemes (other than uniform quantization) that optimally allocate bandwidth to each of the sensors is also another topic of interest. Finally, in-network inference where only the summary of estimated parameters are communicated to the sink, rather than entire observations, will be another framework that one might consider. 

\section*{Acknowledgment}
This work was partially supported by the National Science Foundation under Grant No. $0925854$, the Australian Research Council and the DEST International Science Linkage Grants. The authors would like to thank Sutharshan Rajasegarar, Bharat Sundaram of The University of Melbourne and Hao Chen, Ruixin Niu (both previously at Syracuse University but now at Boise State University and Virginia Commonwealth University respectively) for valuable discussions during the initial stages of this research. The authors would also like to thank the anonymous reviewers for their suggestions which helped improve the original draft of this paper substantially.

\appendices
\section{Proof of Proposition 1} \label{app:prop:1}
In Proposition \ref{pr:matrix:approx}, we need to show equation \eqref{igRi:app:inM}, which is $\left( \bo I+\gamma\bo R \right)^{-1} = \bo I -\nu \bo M+\mathcal O(y^N)$. 
Define $\widetilde{\bo M}$ be such that $\left( \bo I+\gamma\bo R \right)^{-1} = \bo I -\nu \widetilde{\bo M}$, i.e., the exact form of \eqref{igRi:app:inM}. It suffices to prove that $\bo M = \widetilde{\bo M}+\mathcal O(y^N)$, or equivalently
\begin{align}
\widetilde{\bo M}^{-1}\bo M &= \bo I+\mathcal O(y^N) \label{mt:m:is:I}.
\end{align}
To prove \eqref{mt:m:is:I}, we need $\widetilde{\bo M}^{-1}$. Definition of $\widetilde{\bo M}$ implies
\begin{align}
\widetilde{\bo M}^{-1}=\nu (\bo I+ \bo R^{-1}/\gamma). \label{mti:is:Igiri}
\end{align}
The inverse of $\bo R_U$ (given by \eqref{def:RU}) is well known \cite{Niu10}. The inverse of $\bo R$ is similar to that of $\bo R_U$ except for the top-left element. Specifically,
\small
\begin{align}
\mathbf{R}^{-1} &= \begin{bmatrix}
1+\varrho_\tau  & -\rho    & \cdots & 0        & 0      \\
-\rho   & 1+\rho^2 & \ddots & 0        & 0      \\
\vdots  & \ddots   & \ddots & \ddots   & \vdots \\
0       & 0        & \ddots & 1+\rho^2 & -\rho  \\
0       & 0        & \cdots &  -\rho   & 1      \\
\end{bmatrix}.  \label{RC:inv}
\end{align}
\normalsize
From \eqref{mti:is:Igiri} and with the help of some identities that follow from \eqref{def:ynukappa}, namely $\nu(1+1/\gamma)=(1-\rho y+y^2)/(1-y^2)$ and $\nu(1+(1+\rho^2)/\gamma)=(1+y^2)/(1-y^2)$,
we have,
\small
\begin{align}
&\widetilde{\bo M}^{-1} =\frac{1}{1-y^2} \begin{bmatrix}
m_\tau  & -y         & \cdots & 0          & 0                   \\
-y         & 1+y^2  & \ddots & 0          & 0                   \\
\vdots   & \ddots   & \ddots & \ddots   & \vdots            \\
0          & 0          & \ddots & 1+y^2  & -y  \\
0          & 0          & \cdots &  -y        & 1-\rho y+y^2 \\
\end{bmatrix}, \nonumber 
\end{align}
\normalsize
where $m_\tau\triangleq 1-\rho y+y^2+\varrho_\tau y/\rho$. Constructing $\bo M$ from Equation \eqref{def:M:structure}, we can now directly verify \eqref{mt:m:is:I}. Specifically, the residual is of the following structure 
\small
\begin{align}
\widetilde{\bo M}^{-1} \bo M-\bo I =\frac{y^N}{r_0}
\begin{bmatrix}
0        & 0         & \cdots  & r_1       & r_3      \\
0        & 0         & \iddots & r_2       & 0        \\
\vdots & \iddots & \iddots & \iddots  & \vdots \\
r_1     & r_2      & \iddots & 0          & 0        \\
r_2     & 0         & \cdots  & 0          & 0        \\
\end{bmatrix}, \label{mtci:mc:residual}
\end{align}
\normalsize
where $r_0\triangleq (1-\rho y)(\rho-\rho^2y+\varrho_\tau y)$, $r_1\triangleq -\varrho_\tau y$, $r_2\triangleq -\rho(\rho-y)(1-\rho y)+\varrho_\tau (1+y^2-\rho y)$ and $r_3\triangleq r_2+\varrho_\tau y(\varrho_\tau-\rho^2)/\rho$. Since \eqref{mtci:mc:residual} are terms of order $\mathcal O(y^N)$, this completes the proof of \eqref{mt:m:is:I} and hence Proposition \ref{pr:matrix:approx}.

\section{Polynomial Approximation of $\bo X' \bo M \bo X$} \label{app:xtmx}
First, we would decompose $\bo M$ (see \eqref{def:M:structure}) to aid analytical calculations. Note that
\begin{align}
\bo M &= \bo I +\bo M_1+\eta_\tau \bo M_2+\kappa \bo M_3,  \label{M:decompose}
\end{align}
and matrices $\bo M_1,\bo M_2\mbox{ and }\bo M_3$ are defined by
\small
\begin{align}
\bo M_1 &\triangleq  \left[ \begin{array}{ccccc}
0             & y            & \cdots & y^{N-2} & y^{N-1}  \\
y             & 0            & \ddots & y^{N-3} & y^{N-2} \\
\vdots      & \ddots    & \ddots & \ddots     &  \vdots   \\
y^{N-2}  & y^{N-3} & \ddots & 0            &  y          \\
y^{N-1}  & y^{N-2} & \cdots & y            &  0
\end{array} \right], \label{def:M1}
\end{align}
\begin{align}
\bo M_2 &\triangleq  \left[ \begin{array}{cccccc}
1            & y           & \cdots    & y^{N-2} & 0          \\
y            & \iddots   & \iddots   & \iddots   & \vdots   \\
\vdots     & \iddots   & \iddots   & \iddots   & \vdots   \\
y^{N-2} & \iddots   & \iddots   & \iddots   & 0          \\
0            & \cdots    & \cdots    & 0           & 0
\end{array} \right], \mbox{ and} \label{def:M2}
\end{align}
\begin{align}
\bo M_3 &\triangleq  \left[ \begin{array}{cccccc}
0           & 0             & \cdots    & 0            & y^{N-1}    \\
0           & \iddots     & \iddots   & y^{N-1} & y^{N-2}    \\
\vdots    & \iddots     & \iddots   & \iddots   & \vdots        \\
0           & y^{N-1}  & \iddots   & \iddots   & y               \\
y^{N-1}& y^{N-2}  & \cdots    & y            & 1
\end{array} \right], \label{def:M3}
\end{align}
\normalsize
respectively. To calculate $\bo X' \bo M \bo X$, we compute each of terms in \eqref{M:decompose}. $\bo X' \bo X$ is given by \eqref{xtx:kl} and \eqref{def:Bcal:qi}. For $\bo X' \bo M_1 \bo X$, we refer to \eqref{def:M1} and collect the identical powers of $y$ from either side of the principal diagonal,
\small
\begin{align}
&[\bo X' \bo M_1 \bo X]_{k,l}
=\sum_{i=1}^{N-1}y^i\sum_{j=1}^{N-i}(j^k(i+j)^l+j^l(i+j)^k) \nonumber \\
&\stackrel{(a)}{=} \sum_{i=1}^{N-1}y^i\sum_{j=1}^{N-i}\sum_{r=0}^{k+l} A_r i^r j^{k+l-r} \nonumber \\
&= \sum_{i=1}^{N-1}y^i\sum_{r=0}^{k+l} A_r i^r \sum_{j=1}^{N-i} j^{k+l-r} \nonumber \\
&\stackrel{(b)}{=} \sum_{i=1}^{N-1}y^i\sum_{r=0}^{k+l} A_r i^r \sum_{s=0}^{k+l-r} B_{s,r} (N-i)^{k+l-r+1-s} \nonumber \\
&\stackrel{(c)}{=} \sum_{i=1}^{N-1}y^i\sum_{r=0}^{q-1} A_r i^r \sum_{s=0}^{q-r-1} B_{s,r} \sum_{t=0}^{q-r-s}C_{t,s,r}N^{q-r-s-t}(-i)^t \nonumber \\
&\stackrel{(d)}{=} \sum_{i=1}^{N-1}y^i\sum_{r=0}^{q-1} \sum_{s=0}^{q-r-1} \sum_{t=0}^{q-r-s} X_{r,s,t} \nonumber \\
&\stackrel{(e)}{=} \sum_{i=1}^{N-1}y^i\left(\sum_{u=0}^{q-1} \sum_{v=0}^{u} \sum_{r=0}^{v} X_{r,u-v,v-r}+ \sum_{v=1}^{q} \sum_{r=0}^{v-1} X_{r,q-v,v-r}\right)\nonumber \\
&\stackrel{(f)}{=} \sum_{i=1}^{N-1}y^i\left(\sum_{u=0}^{q-1} N^{q-u}\sum_{v=0}^{u} i^v\sum_{r=0}^{v} D_{r,v,u}+ \sum_{v=1}^{q} i^v\sum_{r=0}^{v-1} D_{r,v,q}\right)\nonumber \\
&\stackrel{(g)}{=} \sum_{u=0}^{q-1} N^{q-u}\underbrace{\sum_{v=0}^{u} Y_v\sum_{r=0}^{v} D_{r,v,u}}_{\triangleq \mathcal A_{u,M_1}, \; u\le k+l}+ \underbrace{\sum_{v=1}^{q} Y_v\sum_{r=0}^{v-1} D_{r,v,q}}_{\triangleq \alpha_{M_1}}+\mathcal O(N^q y^N), \label{xt:m1:x}
\end{align}
\normalsize
where in step (a) we collect all like powers of $i$ and $j$ and define $A_r\triangleq \binom{l}{r}+\binom{k}{r}$, (b) follows from the summation formula in \eqref{def:Bcal:qi} and the definition $B_{s,r}\triangleq \mathcal B_{k+l-r,s}$, (c) is due to binomial expansion and definitions $q\triangleq k+l+1$ and  $C_{t,s,r}\triangleq \binom{q-r-s}{t}$. Step (d) follows from the definition $X_{r,s,t}\triangleq i^{r+t} N^{q-r-s-t} A_r B_{s,r} C_{t,s,r} (-1)^t$, step (e) is an identity involving series rearrangement (i.e., true for any $X_{r,s,t}$, by defining $u\triangleq r+s+t$ and $v\triangleq r+t$), step (f) follows from defining $D_{r,v,u}\triangleq A_r B_{u-v,r} C_{v-r,u-v,r}(-1)^{v-r}$. Step (g) follows from the definition $Y_v\triangleq \sum_{i=1}^{\infty} i^v y^i$ (which converges for finite $v$ since $y<1$ and the error term is of the order $N^v y^N$). Quantities of the form $Y_v$ are also known as poly-logarithms. In the last step, we denote the polynomial coefficients by $\mathcal A_{u,M_1}$ and the constant term by $\alpha_{M_1}$.

For $\bo X' \bo M_2 \bo X$, we start with \eqref{def:M2} and collect the identical powers of $y$ from the top-left half,
\small
\begin{align}
[\bo X' &\bo M_2 \bo X]_{k,l} = \sum_{i=1}^{N-1}y^{i-1}\sum_{j=1}^{i} j^k (i+1-j)^l \nonumber \\
&=\underbrace{\sum_{i=1}^{\infty}y^{i-1}\sum_{j=1}^{i} j^k (i+1-j)^l}_{\triangleq \alpha_{M_2}} +\mathcal O(N^q y^N), \label{xt:m2:x}
\end{align}
\normalsize
where $\alpha_{M_2}$ is a constant. Similarly, for $\bo X' \bo M_3 \bo X$, we collect the identical powers of $y$ from the bottom-right half of \eqref{def:M3},
\small
\begin{align}
&[\bo X' \bo M_3 \bo X]_{k,l} 
=\sum_{i=1}^{N-1}y^{i-1}\sum_{j=1}^{i} (N-j+1)^k(N+j-i)^l \nonumber \\
&\stackrel{(a)}{=}\sum_{i=1}^{N-1}y^{i-1}\sum_{j=1}^{i} \sum_{r=0}^k G_r N^{k-r}(-j+1)^r \sum_{s=0}^l H_s N^{l-s}(j-i)^s \nonumber \\
&\stackrel{(b)}{=}\sum_{i=1}^{N-1}y^{i-1}\sum_{j=1}^{i} \sum_{t=0}^{k+l} N^{k+l-t}  \sum_{s=0}^t K_{t,s} \nonumber \\
&\stackrel{(c)}{=}\sum_{t=0}^{k+l} N^{k+l-t} \mathcal A_{t,M_3}+\mathcal O(N^{q}y^N), \label{xt:m3:x}
\end{align}
\normalsize
where (a) follows from binomial expansion and definitions $G_r\triangleq \binom{k}{r}$, $H_s\triangleq \binom{l}{s}$. In (b), we have rearranged the sum to group similar exponents of $N$ with the transformation $t\triangleq r+s$ and defined $K_{t,s}\triangleq G_{t-s}(-j+1)^{t-s} H_s (j-i)^s$, (c) follows from similar arguments while deriving \eqref{xt:m1:x} and \eqref{xt:m2:x} and the definition
$\mathcal A_{t,M_3}\triangleq \sum_{i=1}^{\infty}y^{i-1}\sum_{j=1}^{i} \sum_{s=0}^t K_{t,s}, \; t\le k+l$.

We note that all the constants $\mathcal A_{u,M_1},\alpha_{M_1},\alpha_{M_2},\mathcal A_{t,M_3}$ depend on $k$ and $l$. We can now prove Proposition \ref{pr:xtmx} by composing $\bo X' \bo M \bo X$ with the help of \eqref{xt:m1:x}, \eqref{xt:m2:x} and \eqref{xt:m3:x} in \eqref{M:decompose}. Define $\mathcal A_{-1,M_3}=0$. The constants $\mathcal A_{k,l,i}$ and $\alpha_{k,l}^{(\tau)}$ in Proposition \ref{pr:xtmx} are given by
\begin{align}
\begin{split}
\mathcal A_{k,l,i}&=\mathcal B_{k+l,i}+\mathcal A_{i,M_1}+\kappa\mathcal A_{i-1,M_3}, \; 0\le i \le k+l, \\
\alpha_{k,l}^{(\tau)}&=\alpha_{M_1}+\eta_\tau \alpha_{M_2}+\kappa \mathcal A_{k+l,M_3},
\end{split} \label{def:A:alpha:formula}
\end{align}
respectively. Some of these constants are enumerated in Table \ref{tbl:A:alpha}. This completes the proof. For example, to compute $\mathcal A_{k,l,0}$, note that $\mathcal B_{k+l,0} =\frac{1}{k+l+1}$, $\mathcal A_{0,M_1} =\frac{2Y_0}{k+l+1}$, $Y_0=\frac{y}{1-y}$, hence
\begin{align}
\mathcal A_{k,l,0}=\mathcal B_{k+l,0}+Y_0 \mathcal A_{0,M_1}= \frac{1+y}{1-y}\frac{1}{k+l+1}, 
\end{align}
which agrees with \eqref{def:A:alpha}.

\begin{table*}[!]
\centering
\begin{tabular}{|c|cccc|}
  \hline
   $i\downarrow$	& $\mathcal B_{k+l,i}$ 	&  $\mathcal A_{i,M_1}$  & $\mathcal A_{i-1,M_3}$ & $Y_i$\\
  \hline
   $0$	            & $\frac{1}{k+l+1}$        &   $\frac{2}{k+l+1}Y_0$                                                                                                                                                              &    $0$                                                                                                & $\frac{y}{1-y}$                           \\ 
   $1$	            & $\frac{1}{2}$               &   $Y_0-Y_1$                                                                                                                                                                               &    $\frac{1}{(1-y)^2}$                                                                        & $\frac{y}{(1-y)^2}$                     \\
   $2$                & $\frac{k+l}{12}$           &  $\frac{k+l}{6}Y_0-\frac{k+l}{2}Y_1+\frac{{k\choose 2}+{l\choose 2}}{k+l-1}Y_2$                                                                     &   $-\frac{(k+l)y}{(1-y)^3}$                                                                 & $\frac{y(1+y)}{(1-y)^3}$             \\
   $3$                & $0$                              &  $-\frac{{k+l \choose 2}}{6}Y_1+\frac{{k\choose 2}+{l\choose 2}}{2}Y_2 -\frac{{k\choose 2}+{l\choose 2}-\frac{kl}{2}}{3}Y_3$   &  $\frac{y((1+y)\left({k\choose 2}+{l\choose 2})+k l y\right)}{(1-y)^4}$ & $\frac{y(1+4y+y^2)}{(1-y)^4}$  \\
  \hline
\end{tabular}\\
\vspace{0.1in}
\begin{tabular}{|c|ccc|}
  \hline
   $(k,l)\downarrow$	& $\alpha_{M_1}$ 	&  $\alpha_{M_2}$  & $\mathcal A_{k+l,M_3}$\\
  \hline
   $(0,0)$    & $-\frac{2y}{(1-y)^2}$      &   $\frac{1}{(1-y)^2}$    & $\frac{1}{(1-y)^2}$      \\ 
   $(0,1)$,$(1,0)$    & $-\frac{y}{(1-y)^2}$       &   $\frac{1}{(1-y)^3}$    & $-\frac{y}{(1-y)^3}$     \\
   $(1,1)$    & $\frac{2y^2}{(1-y)^4}$  & $\frac{1}{(1-y)^4}$      &  $\frac{y^2}{(1-y)^4}$  \\
  \hline
\end{tabular}
\caption{Values of several constants helpful to derive $\mathcal A_{k,l,i}$ and $\alpha_{k,l}^{(\tau)}$. } \label{tbl:A:alpha}
\end{table*}

\section{FIM and CRB for $\rho=1$} \label{app:fim:crb:r1}
First we derive the FIM given by Equation \eqref{JC:alpha1}. Let the Taylor Series of the FIM be of the form
\begin{align}
\bo J &=\frac{1}{\sigma^2} \bo E \left[ \bo Q_0+\frac{1}{N}\bo Q_1+\frac{1}{N^2}\bo Q_2+\mathcal O\left(\frac{1}{N^3}\right)\right] \bo E. \label{JC:taylor}
\end{align}
The first term $\bo Q_0$ is equal to $\xi_2 \bo f \bo f'$ (follows from \eqref{J:EH} and the fact that $\xi_0=\xi_1=0$ for $\rho=1$).  The second order term $\bo Q_1$ follows from collecting third-order terms in \eqref{xtx:kl} and \eqref{xtmx:kl} and adding them according to \eqref{J:approx}, i.e.,
\begin{align}
[\bo Q_1]_{k,l}=\left\{ \begin{array}{rl}
0,                                                            & k=l=0, \\
-\nu \alpha_{k,l}^{(\tau)},                        & k+l=1, \\
\mathcal B_{k+l,2}-\nu \mathcal A_{k,l,2}, & k+l\ge 2.
\end{array} \right. \label{G:alpha}
\end{align}
From definitions \eqref{def:ynukappa} and identity \eqref{identity:gy}, we obtain the following simplifications for $\rho=1$, $\nu=\frac{1-y}{1+y}$, $\kappa =y$ and $\frac{1}{\gamma}=\frac{y}{(1-y)^2}$,
which when applied to \eqref{def:A:alpha:formula} (and using values from Table \ref{tbl:A:alpha}) leads to $[\bo Q_1]_{k,l} =\frac{kl}{k+l-1} \frac{1}{\gamma}$ for $k+l\ge 2$,
which explains the term $\xi_4$ in \eqref{def:xi34}. Similarly, $\bo Q_2$ follows from collecting fourth-order terms,
\begin{align}
[\bo Q_2]_{k,l}=\left\{ \begin{array}{rl}
0,                                                            & k+l\le 1, \\
-\nu \alpha_{k,l}^{(\tau)},                        & k+l=2, \\
\mathcal B_{k+l,3}-\nu \mathcal A_{k,l,3}, & k+l\ge 3,
\end{array} \right.
\end{align}
the last term of which when simplified yields $[\bo Q_2]_{k,l}=-k l \frac{y^2}{(1-y)^3}$ for $k+l\ge 3$,
that explains the term $\xi_5$ in \eqref{def:xi34}. 
The derivation of the CRLB in Equation \eqref{crlb:bj:alpha1} follows from the block inversion of $\bo J$ and subsequent application of Lemma \ref{pr:inv:hilbert},
\begin{align}
\begin{bmatrix}
a     & \bo b' \\
\bo b & \bo C
\end{bmatrix}^{-1}&=
\begin{bmatrix}
\frac{1}{a}+\frac{\bo b'\left[\bo C-\frac{1}{a}\bo b\bo b'\right]^{-1}\bo b}{a^2} & * \\
* & \left[\bo C-\frac{1}{a}\bo b\bo b'\right]^{-1}
\end{bmatrix}  \nonumber \\
&\triangleq \begin{bmatrix}
m     & * \\
* & \bo M
\end{bmatrix}, \mbox{ say}. \label{block:inv}
\end{align}
From \eqref{JC:alpha1}, we have $a= \xi_2^{(\tau)}$, $\bo b= \frac{1}{N}\xi_3^{(\tau)}\bo f'+\mathcal O\left(\frac{1}{N^2} \right)$ and
\small
\begin{align*}
\bo C&=\frac{1}{N}\bo D \left[\xi_4 \bo H+\frac{1}{N}\left(\xi_5 \bo e \bo e'+\xi_6^{(\tau)} \bo f \bo f'\right)+\mathcal O\left(\frac{1}{N^2}\right)\right] \bo D,
\end{align*}
\normalsize
where subscript $P$ is dropped from $\bo H_P, \bo e_P$ and $\bo f_P$. Next we apply Lemma \ref{pr:inv:hilbert} to obtain the diagonal elements $[\bo M]_{q,q}$ of  block $\bo M$. By substituting $c_0=\xi_4,\, c_1=\xi_5\, c_2=\xi_6^{(\tau)}-\left(\xi_3^{(\tau)}\right)^2/\xi_2^{(\tau)}$, and observing the facts that $[\bo D^{-1}]_{p,p}=\frac{1}{p}$ and $L_{P-1,p-1}=P^2/p^2$, we obtain $[\bo V]_{p,p}$ for $p\ge 1$ as in \eqref{crlb:bj:alpha1}. For the top-left element, $m$, the second order term is
\small
\begin{align*}
\frac{\bo b'\bo M \bo b}{a^2} &= \frac{\left(\xi_3^{(\tau)}\right)^2 }{\left(\xi_2^{(\tau)}\right)^2} \bo M_{1,1},\; \mbox{ but } \bo M_{1,1}=\frac{K_{P-1,0}}{N \xi_4}+\mathcal O\left(\frac{1}{N^2}\right),
\end{align*}
\normalsize
which completes the derivation of \eqref{crlb:bj:alpha1}.

\bibliographystyle{IEEEtran}
\bibliography{driftbib}

\begin{IEEEbiography} 
[{\includegraphics[width=1in,height=1.25in,clip,keepaspectratio]{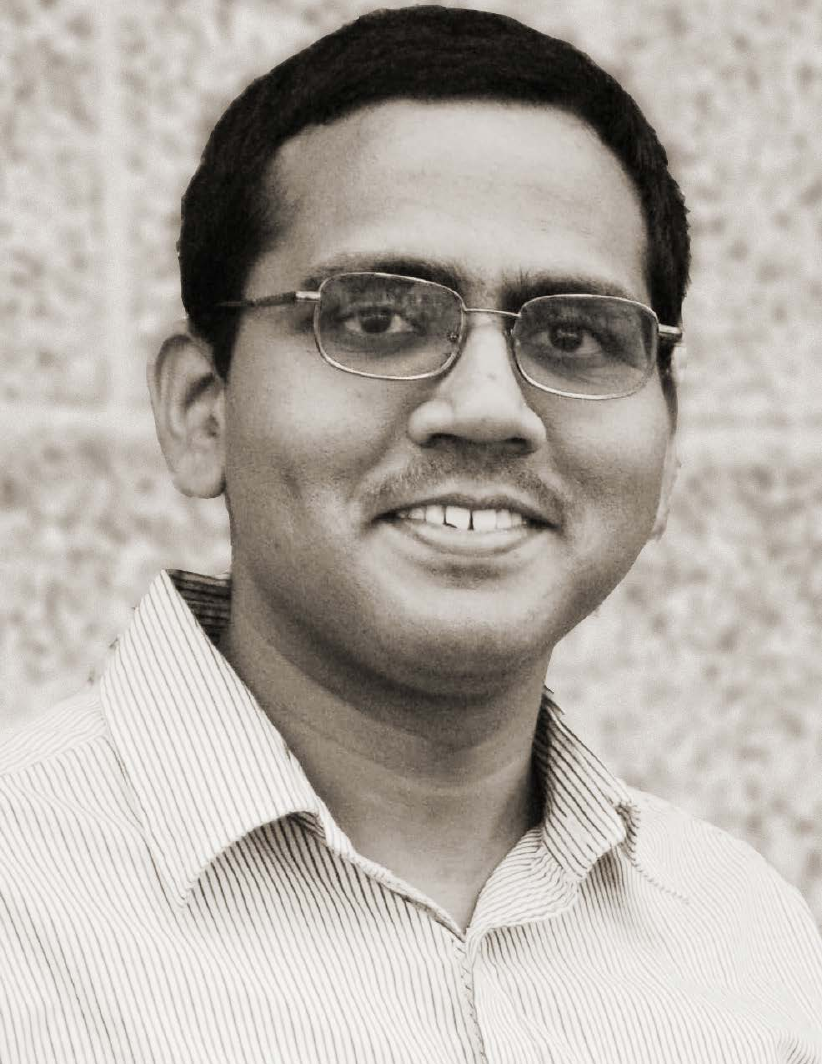}}]{Swarnendu Kar}
(S'07) received the B.Tech. degree in electronics and electrical communication engineering from the Indian Institute of Technology, Kharagpur, India, in 2004 and the M.S. degree in mathematics from Syracuse University, Syracuse, NY, in 2009, where he is currently working toward the Ph.D. degree in electrical engineering.
He was a Video Systems Engineer with Ittiam Systems Pvt. Ltd., Bangalore, India, during 2004--2006. He was a visiting student at The University of Melbourne, Parkville, Australia, during October--December 2009. His research interests include detection and estimation theory and distributed estimation in the context of sensor networks.
\end{IEEEbiography}

\begin{IEEEbiography} 
[{\includegraphics[width=1in,height=1.25in,clip,keepaspectratio]{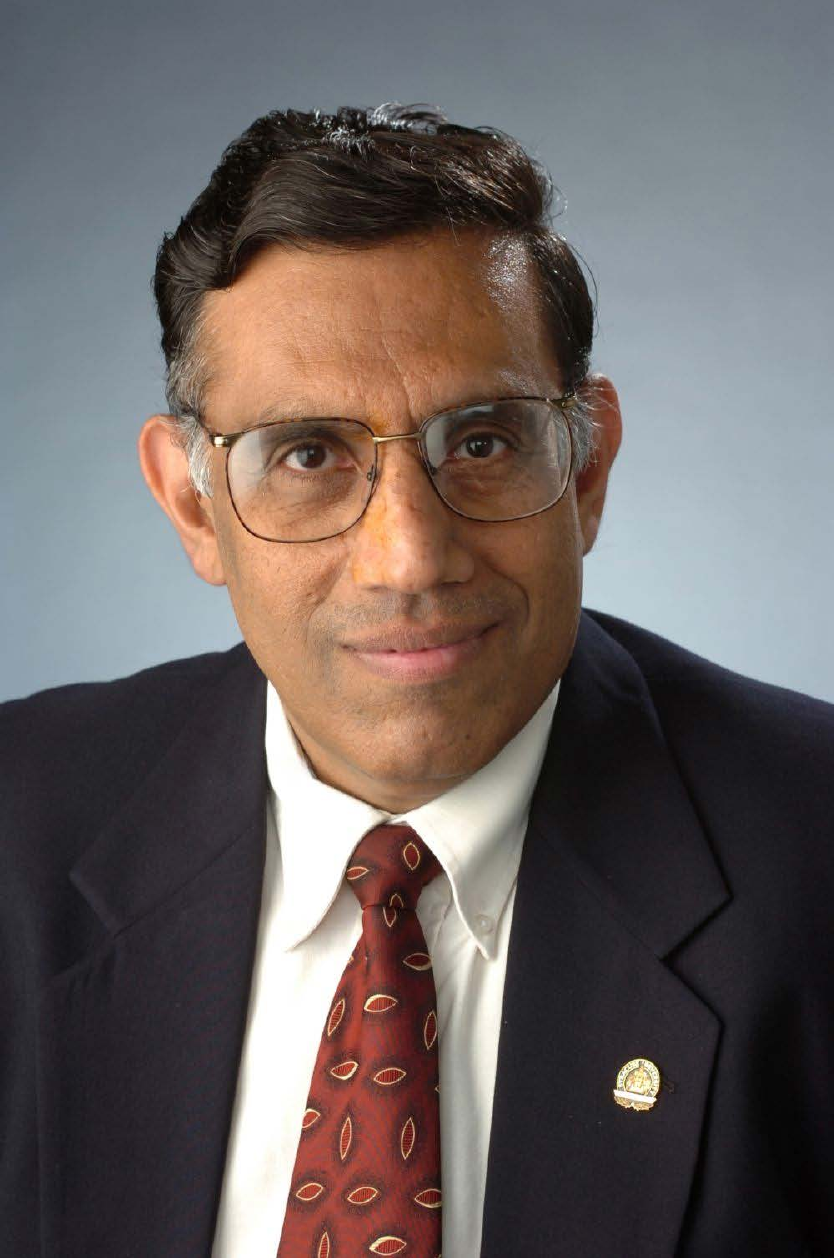}}]{Pramod K. Varshney}
(S'72-–M'77-–SM'82-–F'97) was born in Allahabad, India, on July 1, 1952. He received the B.S. degree in electrical engineering and computer science (with highest hons.), and the M.S. and Ph.D. degrees in electrical engineering from the University of Illinois at Urbana-Champaign in 1972, 1974, and 1976 respectively. From 1972 to 1976, he held teaching and research assistantships at the University of Illinois. Since 1976, he has been with Syracuse University, Syracuse, NY, where he is currently a Distinguished Professor of Electrical Engineering and Computer Science and the Director of CASE: Center for Advanced Systems and Engineering. He served as the Associate Chair of the department from 1993 to 1996. He is also an Adjunct Professor of Radiology at Upstate Medical University, Syracuse, NY. His current research interests are in distributed sensor networks and data fusion, detection and estimation theory, wireless communications, image processing, radar signal processing, and remote sensing. He has published extensively. He is the author of \emph{Distributed Detection and Data Fusion} (Springer-Verlag, 1997). He has served as a consultant to several major companies. Dr.Varshney was a James Scholar, a Bronze Tablet Senior, and a Fellow while at the University of Illinois. He is a member of Tau Beta Pi and is the recipient of the 1981 ASEE Dow Outstanding Young Faculty Award. He was elected to the grade of Fellow of the IEEE in 1997 for his contributions in the area of distributed detection and data fusion. He was the Guest Editor of the Special Issue on Data Fusion of the PROCEEDINGS OF THE IEEE, January 1997. In 2000, he received the Third Millennium Medal from the IEEE and Chancellor’s Citation for exceptional academic achievement at Syracuse University. He is the recipient of the IEEE 2012 Judith A. Resnik Award. He serves as a Distinguished Lecturer for the IEEE Aerospace and Electronic Systems (AES) Society. He is on the Editorial Board of the \emph{Journal on Advances in Information Fusion}. He was the President of International Society of Information Fusion during 2001.
\end{IEEEbiography}

\begin{IEEEbiography}[{\includegraphics[width=1in,height=1.25in,clip,keepaspectratio]{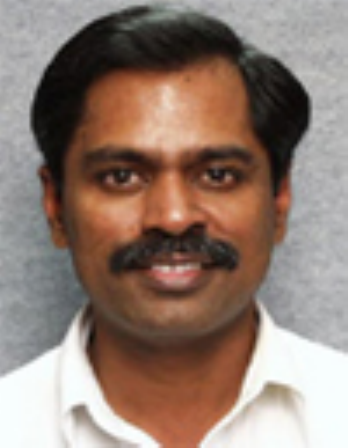}}]{Marimuthu Palaniswami}
received his ME from the Indian Institute of Science, India, MEngSc from the University of Melbourne and PhD from the University of Newcastle, Australia before rejoining the University of Melbourne. He has published over 350 refereed research papers in highly reputed journals and conferences. He currently leads a large  ARC Research Network on Intelligent Sensors, Sensor Networks, and Information Processing (ISSNIP), which is a network centre of excellence with complementary funding for fundamental research, test beds, international linkages, and industry linkages. Previously, he was a director of Centre of Networked Decision Systems (CENDS). His leadership includes external reviewer to an International Research Centre, selection panel member for senior appointments/promotions, grants panel member for NSF (USA), advisory board member for European FP6 grant centre, steering committee member for NCRIS GBROOS and SEMAT, and board member for IT and SCADA companies. His academic excellence is recognized by several invited presentations of plenary/keynote lectures and serving as General Chair, he organised over 10 international conferences.. He has served as an associate editor for journals/transactions including IEEE Transactions on Neural Networks, International Journal Distributed sensor Networks, International Journal of Biomedical Engineering. He was given a Foreign Specialist Award by the Ministry of Education, Japan.   In recognition of his community and industry engagement, he received several Uni of Melb Knowledge Transfer Excellence Awards. His research interests include SVMs, sensors and sensor networks, machine learning, neural network, pattern recognition, signal processing, and control. The interdisciplinary themes for his  large scale projects cover Biology, Medicine, Mathematics, Architecture, and Engineering. He is a fellow of the IEEE.

\end{IEEEbiography}

\end{document}